\documentclass[prb,twocolumn,amsmath,groupedaddress,amssymb,floatfix,footinbib,bibnotes,longbibliography,superscriptaddress]{revtex4-2}

\usepackage{amsmath,amsfonts,amssymb}
\usepackage{mathrsfs}
\usepackage{graphicx}
\usepackage[english]{babel} 
\usepackage{verbatim}
\usepackage[colorlinks=true,citecolor=blue,linkcolor=magenta]{hyperref}
\usepackage[usenames, dvipsnames]{color}
\usepackage{bbold}
\usepackage{chemformula}

\newcommand{\UCB}{Department of Physics, University of California, Berkeley, CA 94720, USA}
\newcommand{\LBL}{Materials Sciences Division, Lawrence Berkeley National Laboratory, Berkeley, California, 94720, USA}

\begin{document}
\title{Dynamics of fractionalized mean-field theories: consequences for Kitaev materials }

\author{Tessa Cookmeyer}
\email[]{tcookmeyer@berkeley.edu}
\affiliation{\UCB}
\affiliation{\LBL}

\author{Joel E. Moore}
\affiliation{\UCB}
\affiliation{\LBL}

\begin{abstract}
    There have been substantial recent efforts, both experimentally and theoretically, to find a material realization of the Kitaev spin-liquid--the ground state of the exactly solvable Kitaev model on the honeycomb lattice. Candidate materials are now plentiful, but the presence of non-Kitaev terms makes comparison between theory and experiment challenging.
    We rederive time-dependent Majorana mean-field theory and extend it to include quantum phase information, allowing the direct computation of the experimentally relevant dynamical spin-spin correlator, which  reproduces exact results for the unperturbed model. In contrast to previous work, we find that small perturbations do not substantially alter the exact result, implying that $\alpha$-\ch{RuCl3} is perhaps farther from the Kitaev phase than originally thought. Our approach generalizes to any correlator and to any model where Majorana mean-field theory is a valid starting point.
\end{abstract}

\maketitle
\section{Introduction}

 The Kitaev model describes spin-1/2's on the honeycomb lattice with a bond-dependent Ising interaction \cite{kitaev2006}. Remarkably, it is exactly solvable by a transformation to Majorana fermions due to the appearence of an extensive number of conserved quantities. The ground state has the fascinating property that in a weak magnetic field the low-energy excitations are non-Abelian anyons \cite{kitaev2006}; beyond the intrinsic interest, these anyons could form the basis for a topological quantum memory device \cite{Nayak2008}.
 
 While the Kitaev model was first introduced without a clear path towards material realization, Jackeli and Khaliluin discovered one such route in 4$d$/5$d$ transition metals \cite{Jackeli2009}. An alternative pathway involving the 3$d$ transition metal Co has recently been discovered \cite{Liu2018,Liu2020,Sano2018}, and there are now several candidate materials for realizing Kitaev physics \cite{winter2017models,takagi2019} such as \ch{Na2IrO3} \cite{Ye2012, Comin2012, hwan2015direct,Singh2010,Singh2012,Choi2012,Liu2011}, \ch{Li2IrO3} \cite{Singh2012, Williams2016,Biffin2014, Winter2016}, \ch{H3LiIr2O6} \cite{kitagawa2018,takagi2019}, \ch{Na2Co2TeO6} \cite{lin2021field}, and $\alpha$-\ch{RuCl3} \cite{banerjee2018,Banerjee2017,banerjee2016,Ran2017,nasu2016}. The ``smoking-gun'' evidence of a quantized Thermal Hall effect has been found in $\alpha$-RuCl${}_3$  \cite{kasahara2018, yokoi2021, bruin2022}, though sample-dependence has complicated efforts to reproduce the result \cite{yamashita2020,czajka2022,lefranccois2021}. 
 
 Due to the convenience of an exact solution, the Kitaev model without additional terms is often used to compare against experiments, for instance in inelastic neutron scattering \cite{Banerjee2017,banerjee2016} and  thermal Hall effect \cite{Kasahara2018Unusual} experiments.  In the candidate materials, however, the microscopic spin Hamiltonian contains non-Kitaev terms \cite{Winter2016,Slagle2018,takagi2019} such as Heisenberg and ``$\Gamma$'' terms.   It is therefore important to have a general method to compute static and dynamic quantities near the pure-Kitaev model point and to know how such terms modify the exact results.

 Standard methods such as (infinite) density-matrix renormalization group \cite{Gohlke2018,Gohlke2017,McClarty2018}, (non-)linear spin-wave theory \cite{Chern2021,McClarty2018, Cookmeyer2018,Yamaji2016, Janssen2020, Winter2018, Zhang2021swt, Gao2021, banerjee2016, Ran2017, Joshi2018, koyama2021}, variational Monte-Carlo \cite{Zhang2021_VMC}, quantum Monte-Carlo \cite{Yoshitake2020,Sato2021}, Monte-Carlo cluster perturbation theory \cite{Ran2022}, Landau-Lifshitz dynamics \cite{samarakoon2022}, and exact diagonalization \cite{Kaib2019,hickey2019,laurell2020,Winter2018} have been used to approach this problem. Although the existence of the exact solution allows some techniques to be more powerful \cite{Zhang2021_VMC,Yoshitake2020,Sato2021}, there are numerous challenges in applying them to a two-dimensional (2D) quantum mechanical system. Instead, one of the most intuitive starting point for taking advantage of and extending the exact result is mean-field theory (MFT) as the conserved quantities in the original model can be thought of as mean fields. Many papers have used MFT in analyzing the Kitaev model with various perturbations \cite{Schaffer2012, Nasu2018static, li2022, Freitas2022, Berke2020, Jiang2020, Choi2018, Zhang2021, Okamoto2013, Seifert2018, Ralko2020, Gao2019, Go2019, Mei2012}, but the authors of Ref.~\onlinecite{Knolle2018} argue that an \textit{augmented} MFT is necessary to correctly compute both static and dynamic quantities at the pure-Kitaev point, which then must be the correct starting point for an extension. It is not clear, however, how to extend their approach to perturbations that mix the itinerant and localized Majoranas, such as a magnetic field, since they are treated distinctly. 
 
 Fundamental to the argument of Ref.~\onlinecite{Knolle2018}, though, is a particular understanding of time-evolution in mean-field theory; namely, time-evolution occurs under the mean-field decoupled Hamiltonian. Although this perspective is commonplace (for example, Refs.~\onlinecite{Liu2019, Halimeh2019, Messio2010, gao2020}), an alternative approach would be time-dependent mean-field theory (TDMFT), as we describe below. TDMFT as applied to electrons has been around, under the name time-dependent Hartree-Fock approximation (TDHFA), since Dirac \cite{dirac1930,Kulander1987,Bonche1976}, and, more recently, has been used to study lattice Hamiltonians relevant to solids \cite{terai1993,tanaka2010,hirano2000}. Working by analogy, the authors of Refs.~\onlinecite{Nasu2019, Minakawa2020, Taguchi2021} extended TDMFT to Majorana fermions and applied it to the Kitaev model in a magnetic field to study quantum quenches \cite{Nasu2019} as well as spin transport \cite{Minakawa2020,Taguchi2021}. Those studies were centered around the computation of expectation values, and therefore the phase of the wave-function was not necessary and not determined. Remarkably, TDMFT, as we will show, is enough to capture all static and dynamic ground-state quantities exactly for the Kiteav model, implying that TDMFT might be integral to understanding time-evolution within mean-field theory in a variety of systems. 
 
 In this paper, we rigorously rederive TDMFT for Majorana's  and provide an explicit expression for the wave-function at time $t$. We then demonstrate how this formalism allows us to compute dynamical quantities in the perturbed Kitaev model that agree with exact results at the Kitaev point, and, as our main result, we find the features of the exact result are more robust than implied by previous work \cite{Knolle2018,Song2016}. Our example quantity is the dynamical spin-spin correlator, $S(\pmb q,\omega)$, but we emphasize that this approach is fully general and should work for any ground-state correlator. Additionally, this approach is not limited to the Kitaev model but instead can be applied whenever Majorana mean-field theory (or a quadratic Majorana Hamiltonian) is a good starting point, and this approach should be generalizable and applicable to bosonic  mean-field theories where the boson number is not conserved.
 
 In Sec.~\ref{sec:TDMFT}, we derive TDMFT for Majoranas. In Sec.~\ref{sec:dynscinKitaev}, we we apply TDMFT to compute the dynamic spin-spin correlator (or dynamic structure factor) in the Kitaev model in the absence and presence of a magnetic field. In Sec.~\ref{sec:results}, we present the results of numerical calculations. We discuss the implications for the results in Sec.~\ref{sec:disc}, and conclude in Sec.~\ref{sec:conc}. 
 
\section{General theory for time-dependent Majorana Mean-field theory}\label{sec:TDMFT}

Our goal in this section is to explain how to perform time-evolution within Majorana mean-field theory. This method should be easily generalizable to arbitrary non-interacting particles, however.  

We will first describe TDHFA, which, in more modern language, is equivalent to a time-dependent mean-field theory decoupling. The analysis is natural and straightforward. For $N$ particles with creation operators $f_i^\dagger$, one computes the self-consistent decoupling of the Hamiltonian and diagonalizes the system into $H(\Theta)=H_0 = \vec f^\dagger M_0(\Theta) \vec f = \sum_n \epsilon_n \gamma_n^\dagger \gamma_n$ via $\vec f = U \vec \gamma$ where $\Theta$ denotes some mean-field parameters like the density $\langle f_i^\dagger f_i\rangle$, and $\epsilon_n\le \epsilon_{n+1}$. The ground state wave-function is given by
\begin{equation}
    |\Psi(t=0)\rangle = \gamma_1^\dagger \gamma_2^\dagger \cdots \gamma_N^\dagger |0\rangle
\end{equation}
with $|0\rangle$ being the vacuum.

One can then imagine evolving this state under some time-dependent Hamiltonian, $H(\Theta(t))=\sum_{n,m} f_m^\dagger M_{m,n}(\Theta) f_n$ which depends on the time-dependent values of $\Theta(t)$, and time-evolution over a short time is given by $e^{-iH(\Theta(t))\Delta t}$. Evolution then follows by commuting the infinitesimal time evolution past each of the $\gamma_i^\dagger$
\begin{equation}
\begin{aligned}
    |\Psi(t+\Delta t)\rangle &= e^{-iH(t) \Delta t} \gamma_1^\dagger(-t) \cdots \gamma_{N}^\dagger(-t) |0\rangle 
    \\&=   \gamma_1^\dagger(-t-\Delta t) \cdots \gamma_{N}^\dagger(-t-\Delta t) |0\rangle
\end{aligned}
\end{equation}
where $\gamma_i^\dagger(-t-\Delta t)=e^{-iH(t)\Delta t}\gamma_i^\dagger(-t) e^{iH(t)\Delta t} = f_j^\dagger U_{ji}(t+\Delta t)$.

 We can compute that  $U(t+\Delta t) = e^{-iM(\Theta)\Delta t}U(t)$ and therefore the columns of $U(t)$ satisfy a Schrodinger equation evolving under the single-particle Hamiltonian $M_{n,m}(\Theta)$. It is then straightforward to compute any expectation needed for $\Theta(t)$ by converting to the basis of $\gamma_i^\dagger(-t)$. In practice, $\gamma_i^\dagger(-t)$ is used to compute $\Theta(t)$, which is used to evolve $\gamma_i^\dagger(-t)$ to $\gamma_i^\dagger(-t-\Delta t)$, though methods with higher order error in $\Delta t$ exist \cite{tanaka2010,Kulander1987}.   

In order to study the Kitaev model, this method has recently been extended to Majoranas \cite{Nasu2019, Minakawa2020,Taguchi2021}. In that case, number is not a conserved quantity, but the authors of Ref.~\onlinecite{Nasu2019} argue by analogy that the same method would work. Here we rigorously derive \textit{why} this analogy holds and provide an explicit expression for the wave function at time $t$.

In the Majorana case, we have some Hamiltonian
\begin{equation}
    H(M^{(t;\theta_{ij})}) = \frac{1}{4} \sum_{ij}c_i M_{ij}^{(t;\theta_{ij})}c_j
\end{equation}
where $M_{ij}$ is a function of time and MFT parameters $\theta_{ij}$ and $c_i^2 = 1$ is a typical Majorana operator. Here $\theta_{ij} = i\langle c_i c_j\rangle$ and is implicitly a function of time. We imagine that any constant term (which can depend on $t$ or $\theta_{ij}$) has been written separately from the Hamiltonian, and that we have $M^T=-M$. The factor of $1/4$ is chosen such that
\begin{equation}\label{eq:Hcomm}
    [H(M),H(N)] = H([M,N])
\end{equation}
as can easily be checked \cite{kitaev2006}. We, at this point, introduce rescaled Majoranas $c_i \to \tilde c_i \sqrt{2}$ so that $\tilde c_i^2 = \frac12$ and $\{ \tilde c_i, \tilde c_j\} = \delta_{ij}$. It is still true that $\tilde c_i^\dagger = \tilde c_i$, and we choose this rescaling because it makes $M$ diagonalizable by a unitary matrix into a complex fermion basis.

At time $t=0$, we diagonalize $H_0 = \frac12\bar a^\dagger \Lambda_0 \bar a$ where $\tilde c_i = U_{0,ij}\bar a_j$ for $\bar a^T = (a_1, a_2, ..., a_N, a^\dagger_1, ..., a_N^\dagger)$ and $\Lambda_0 = \text{diag}\{E_1, E_2, ... ,E_N, -E_1, ..., -E_N\}$. The ground state is now given by the unique state $|v\rangle$ such that $a_i |v\rangle =0$. Arguing by analogy, we should expect that the time-evolved state will always be the vacuum of operators $\bar a^{(t)}=U(t)^\dagger \vec {\tilde c}$ where instantaneously, we evolve the columns of the matrix $U(t)$ via a Schrodinger equation. Noting that infinitesimal time evolution is governed by the quadratic Hamiltonian $H(M^{(t;\theta_{ij})})$, it is clear that 
\begin{equation}
\begin{aligned}
    \bar a^{(t+\Delta t)} &= e^{-iH(M^{(t;\theta_{ij})})\Delta t} \bar a^{(t)} e^{iH(M^{(t;\theta_{ij})})\Delta t} \\&= U(t)^\dagger e^{i\Delta t M^{(t;\theta_{ij})}} \vec{\tilde c}
\end{aligned}
\end{equation} will annhilate $|v_{t+\Delta t}\rangle = e^{-iH(M^{(t;\theta_{ij})})\Delta t}|v_{t}\rangle$ where $|v_t\rangle$ is the vacuum for $\bar a^{(t)}$. 
It follows that  $U(t+\Delta t) = e^{-i\Delta t M^{(t;\theta_{ij})}}U(t)$ implying, once again, that $U(t)$ satisfies a Schrodinger equation under the single-particle matrix $M^{(t;\theta_{ij})}$ confirming our expectation.

However, this calculation does not fix the phase, and it will be necessary in our case. Using standard results for the expression of the relationship between the vacuum states for two different fermionic bases, and the result of Ref.~\onlinecite{Knolle2015} for the evaluation of $\langle e^{-i H(M^{(t;\theta_{ij})})\Delta t}\rangle$, we find
\begin{equation}\label{eq:vac}
  e^{-iH(\mathcal M_t)}|v\rangle = \sqrt{\det X} e^{\frac12 (\vec a^\dagger)^T F  \vec a^\dagger  }|v\rangle .
\end{equation}
The matrix $e^{-i H(\mathcal M_t)} = \prod_n e^{-iH(M^{(t_n,\theta_{ij})})\Delta t}$ is the approximate time evolution operator, and we use the notation $e^{-i \mathcal M_t} = \prod_ne^{-i M^{(t_n,\theta_{ij})}\Delta t}$. The matrices $F = -X^{-1}Y$, $X$, and $Y$ are determined by the change of basis formula between the operators $\bar a^{(t)}$ and $\bar a$, namely 
\begin{equation}\label{eq:bararef}
   \bar a^{(t)} = U^\dagger (t) \vec {\tilde c} = U_0^\dagger e^{i\mathcal M_t}U_0 \bar a =  \begin{pmatrix} X & Y \\
    Y^* & X^* \end{pmatrix}\bar a.  
\end{equation}
 As in Ref.~\onlinecite{Knolle2015}, we evaluate $\sqrt{\text{det} X} = \sqrt{|\text{det} X|} e^{-i\phi(t)/2}$ and the sign ambiguity due to $\phi(t) = \text{arg} [\det(X)]$ is avoided by requiring that $\phi(t)/2$ is a continuous function. 

Now, evolving $|v\rangle$ proceeds as in the number-conserving case. At any time step, we compute $\theta_{ij}$ by rewriting $c_ic_j$ in the $a^{(t)}$ basis and using Eq.~\eqref{eq:vac}. The $\theta_{ij}$ specify the approximate infinitesmial time evolution operator $\mathcal U(t+\Delta t,t) = e^{-iH(M^{(t,\theta_{ij})})\Delta t}$, which is then used to find the $a^{(t+\Delta t)}$ basis and contribution to the phase $\phi(t+\Delta t)$. This procedure can straightforwardly be extended to other states beyond $|v\rangle$, an example of which we will see below. 

An alternative perspective on the above results comes from considering more carefully the approximate time-evolution operator.
\begin{equation}\label{eq:BCH}
\begin{aligned}
    \mathcal U(t,0) &= e^{-i H(\mathcal M_t)}= \prod_{n} e^{-iH(M^{(t_n,\theta_{ij})})\Delta t} 
    \\&= \exp\left[H\left(\log\left(\prod_n e^{-i\Delta t M^{(t_n;\theta_{ij})}}\right)\right)\right] \\
    &=\exp\left[H\left(\log\left(e^{-i\mathcal M_t}\right)\right)\right]
\end{aligned}
\end{equation}
where the second step follows by the Baker-Campbell-Hausdorff theorem since the $H(M)$ distributes over addition, multiplication, and commutation, and $[M,N]$ is still an antisymmetric matrix with no trace \cite{udagawa2021}. This calculation justifies our use of the notation $e^{-i\mathcal M_t}$ from earlier. It is only, therefore, necessary to be able to compute the $\theta_{ij}$ and, instead of evolving the wavefunction, one can just consider updating the time-evolution operator. 

To close this section, we wrap up with a question about the validity of TDMFT. We are making the mean-field approximation because we cannot solve the model exactly--whether or not this approximation is a good starting point depends on the model. Assuming that it is a good starting point, if we wish to compute 
$U(t,0)|\Psi\rangle = e^{-i\mathcal Ht}|\Psi\rangle$ where $\mathcal H$ is any Hamiltonian and $\Psi$ is any state in the Hilbert space, we need to mean-field decouple $\mathcal H$ in some way. If $|\Psi\rangle$ is somehow related to the ground state, one might expect that replacing $\mathcal H$ with $\mathcal H_\text{MF}$, with mean-field parameters determined from the ground state, is the way forward. However, since $|\Psi\rangle$ is not the ground state, we can decouple $\mathcal H$ again at time $t$ with respect to $|\Psi(t)\rangle$, as in the TDMFT introduced above. We show a comparison between these two approaches in Appendix~\ref{app:tdmftvother}, and it is clear that TDMFT captures more of the relevant physics. Since, as we will show, TDMFT reproduces the exact results of the Kitaev model in the absence of perturbations without any kind of tuning, we expect that it will remain a good approximation for small perturbations and finite times. 
To support this expectation, we compare TDMFT directly to density-matrix renormalization group methods in Appendix~\ref{app:compdmrg}, and we find that it is able to qualitatively (and sometimes quantitatively) capture the effect of perturbations.

\section{Dynamical spin-correlators in the Kitaev Model}\label{sec:dynscinKitaev}

We now turn our focus to the Kitaev-Heisenberg-$\Gamma$ model near the Kitaev point in a small magnetic field, $\pmb h = - g\mu_B\mu_0\pmb H$ 
\begin{equation}
\begin{aligned}
H = -  \sum_{\langle i j\rangle_\alpha}\bigg[ &K S_i^\alpha S_j^\alpha +\Gamma \sum_{\beta \ne \bar \beta \ne \alpha} S_i^\beta S_j^{\bar \beta}  
\\+ &J \pmb S_{i}\cdot \pmb S_j\bigg] + \sum_{i} \pmb h \cdot \pmb S_i.
\end{aligned}
\end{equation}
The sum is over all nearest-neighbor bonds and each bond has an index $\alpha=x,y,z$ according to its type.
By substituting $S^\alpha_i = \frac12 i c_i b_i^\alpha$ \cite{kitaev2006} we get
\begin{equation}
\begin{aligned}
    H &= \frac14 \sum_{\langle i j\rangle_{\alpha}}\bigg[K ic_i c_j (ib_i^\alpha b_j^\alpha) + J\sum_{\beta }ic_i c_j (ib_i^\beta  b_j^\beta)\\
    &+\Gamma \sum_{\beta \ne \bar \beta \ne \alpha} ic_ic_j (i b_i^{\bar \beta} b_j^{\beta})\bigg] + \frac12\sum_i \sum_\alpha h_\alpha i c_i b_i^\alpha
\end{aligned}
\end{equation}

If we set $h_\alpha = J=\Gamma=0$, this model can be exactly solved since all the operators $u_{\langle ij\rangle_\alpha} = ib_i^\alpha b_j^\alpha$ commute with $H$ and with each other \cite{kitaev2006}. The ground state is found in the sector with uniform $u_{\langle ij\rangle_{\alpha}}$, and the resulting Hamiltonian is quadratic in the $c_i$.

Beyond an exact expression for the ground state, any dynamic quantity, such as the dyanmic spin-spin  \cite{Baskaran2007,Knolle2015} and dynamic energy current-energy current correlators  \cite{Pidatella2019,Nasu2017}, can be computed exactly. We focus on the former defined as 
\begin{equation}
    S^{\alpha\beta}(\pmb q,\omega) =\frac{1}{N} \sum_{i,j} e^{i\pmb q \cdot (\pmb x_i - \pmb x_j)} \int_{-\infty}^\infty dt e^{-i\omega t} \langle S_i^\alpha(t)S_{j}^\beta(0)\rangle.
\end{equation}
Evaluating the dynamic spin-spin correlator expressions for the Kitaev model is similar in nature to the x-ray mobility edge problem, and multiple exact approaches were derived in Ref.~\onlinecite{Knolle2015}. 

\subsection{Zero-field approach}
We will start by assuming $\pmb h=0$ for simplification and to compare with Ref.~\onlinecite{Knolle2018}. In this case, we can mean-field decouple the Hamiltonian to get $H\approx H_\text{MF} = H_c + H_b +H_C$
\begin{widetext} 
\begin{equation}\label{MFdecoupling}
\begin{aligned}
    H_c &= \frac14 \sum_{\langle i j\rangle_\alpha} \left[(K+J) \langle ib_i^\alpha b_j^\alpha\rangle + J\sum_{\beta \ne \alpha}  \langle ib_i^\beta b_j^\beta\rangle +\Gamma \sum_{\bar \beta \ne \beta \ne \alpha} \langle i b_i^{\bar \beta} b_j^\beta \rangle \right] i c_i c_j=\frac18 \sum_{i,j} M_{ij}^{c} c_i c_j \\
      H_b &= \frac{1}{4} \sum_{\langle i j\rangle_\gamma} \langle ic_i c_j\rangle \left[\Gamma \sum_{\bar \beta \ne \beta \ne \gamma} ib_i^{\bar \beta} b_j^\beta +\sum_{\alpha} \left(K \delta_{\alpha, \gamma} + J \right) ib_i^\alpha b_j^\alpha \right] = \frac18 \sum_{i,j} M_{ij}^{b\alpha} b_i^\alpha b_j^\alpha  \\
      H_C &= -\langle H_c\rangle  =  -
    \frac14 \sum_{\langle i j\rangle_\alpha} \left[(K+J) \langle ib_i^\alpha b_j^\alpha\rangle + J\sum_{\beta \ne \alpha}  \langle ib_i^\beta b_j^\beta\rangle +\Gamma \sum_{\bar \beta \ne \beta \ne \alpha} \langle i b_i^{\bar \beta} b_j^\beta \rangle \right] \langle i c_i c_j\rangle 
\end{aligned}
\end{equation}
\end{widetext}

We will use TDMFT so the expectation values have time dependence. As a convention, we will choose e.g. $H_c^0$ to denote that the expectation values are computed in the ground state, $|v\rangle$. The mean-field expectation values in the ground state are determmined self-consistently using the unperturbed Kitaev model as an initial guess.

We will focus on the dynamic spin-spin correlation, but this approach should work for any correlator. Letting $E_\text{MF}$ be the ground state energy from mean-field theory, we have
\begin{equation}
    S^{\alpha \beta}_{ij}(t) = \langle S^\alpha_i(t) S^\beta_j\rangle  = -\frac14 e^{iE_\text{MF}t} \langle c_i b_i^\alpha U(t,0) c_j b_j^\beta \rangle.
\end{equation}
If we use the above formalism to evolve $|\Psi \rangle = c_j b_j^\beta |v\rangle $ in time (to compute $|\Psi(t)\rangle = \mathcal U(t,0)|\Psi\rangle  \approx U(t,0)|\Psi\rangle$), we can approximate the time evolution operator as $e^{-iH(\mathcal M_t)}$ which implicitly depends on the history of mean-field parameters. Additionally, $\mathcal M_t$ will be block diagonal in the $c$ and $b$, so we can separate the ground state into a tensor product of the ground states of the $c$'s and $b$'s, i.e. $|v\rangle = |v_c\rangle \otimes |v_b\rangle$, and 
\begin{equation}
\begin{aligned}
  \mathcal U(t,0)|\Psi\rangle &= e^{-i\psi(t)}e^{-H_c(\mathcal M_t^c)}e^{-H_b(\mathcal M_t^{b})}
    \\
    e^{-iH_x(\mathcal M^x_t)} &= \prod_n e^{-i \Delta t H_x(t_n)}; \qquad \psi(t) = \int_0^t ds H_C(s)
\end{aligned}
\end{equation} where $H_x(t_n)$ are determined from Eq.~\eqref{MFdecoupling} with time-dependent expectation values. Therefore 
\begin{equation}\label{eq:SQTDMFT}
\begin{aligned}
    S_{ij}^{\alpha \beta}(t) &=\frac14 \delta_{\alpha \beta} e^{iE_\text{MF}t - i\psi(t)}\\
    \times &\langle v_c| c_i e^{-i H_{c}(\mathcal M_t^c)} c_j  |v_c\rangle
    \langle v_b | b_i^\alpha e^{-iH_{b}^\alpha(\mathcal M_t^{b})}b_j^\beta  |v_b \rangle 
\end{aligned}
\end{equation}

In order to evaluate the above expressions, we can use the result of Ref.~\onlinecite{Knolle2015}, which we rederive from Eq.~\eqref{eq:vac} in Appendix~\ref{app:Pfaf}. Additionally, because we need to compute expectation values with respect to $|\Psi(t)\rangle$, we will need to compute correlations like
    $i\langle c_i c_j(t) c_k(t) c_i \rangle$
which follow from a straightforward application of Wick's theorem. 

\subsection{Recovering the exact solution}

At the exactly solvable point $J=\Gamma=\pmb h=0$, it is clear that the three flavors of $b$'s decouple and $H_b = \sum_\alpha H_b^\alpha$ can be diagonalized by the transformation $ib_i^\alpha b_j^\alpha = 1-2\chi_{\langle ij\rangle_\alpha}^\dagger \chi_{\langle ij\rangle_\alpha}$; put another way, $H_b^\alpha$ are all diagonal in the bond-fermion basis \cite{Baskaran2007}. Because we have diagonalized the Hamiltonian and we choose the gauge where $\chi_{\langle ij\rangle_\alpha}^\dagger \chi_{\langle ij\rangle_\alpha} = 0$, it is easy to compute that $\langle i b_k^\beta b_l^\beta\rangle(t) = 1$ if $k$ and $l$ are connected via a $\beta$ bond except that $\langle i b_k^\alpha b_l^\alpha\rangle(t)=-1$ when we are computing $S^{\epsilon \alpha}_{kl}$ (every other expectation is 0 except the trivial $\langle i b_i^\alpha b_i^\alpha\rangle=i$).

Because $H_b^\alpha(t_n)$ is diagonal in the bond-fermion basis, it is clear the bond-fermions cannot move. Breaking the ground state into a product of the ground states of each of the $b^\alpha$'s we therefore compute
\begin{equation}
\begin{aligned}
    \langle v_b | &b_i^\epsilon e^{-iH_b(M_t^{b})}b_j^\alpha |v_b \rangle = \delta_{\alpha\epsilon}\langle v_{b^\alpha} b_i^\alpha e^{-iH_{b}^\alpha(\mathcal M_t^{b^\alpha})}b_j^\alpha |v_{b^\alpha}\rangle \\&\times \langle v_{b^\beta}| e^{-iH_{b}^\beta(\mathcal M_t^{b^\beta})} |v_{b^\beta}\rangle \langle v_{b^\gamma} | e^{-iH_{b}^\gamma(\mathcal M_t^{b^\gamma})} |v_{b^\gamma}\rangle \\
    & = -i \delta_{\alpha \epsilon} e^{i\psi(t)}\langle v_{b^\alpha}| i b_i^\alpha b_j^\alpha|v_{b^\alpha}\rangle.
    \end{aligned}
\end{equation} 
The phase exactly cancels that accumulated from the $H_C$ term because $|\Psi(t)\rangle$ is still an eigenstate of the bond-fermion operators so $ib_i^\alpha b_j^\alpha = \langle i b_i^\alpha b_j^\alpha\rangle$. In the ground state, $\langle b_i^\alpha b_j^\alpha\rangle = 0$ unless $i,j$ are connected by an $\alpha$ bond.

Putting everything together, and noting that $E_\text{MF}$ is exactly the ground state energy for the Kitaev model, we find that we recover the exact result \cite{Baskaran2007,Knolle2015}:
\begin{equation}
\begin{aligned}
H_c(t_n) = H_F &=-\frac12 K i c_i c_j +  \frac{1}{4}\sum_{\langle k l\rangle_\alpha} K i c_k c_l \\
    S_{ij}^{\alpha \alpha}&=-\frac{i}4e^{iE_0t}\langle c_i e^{-iH_F t} c_j \rangle \\
     S_{ii}^{\alpha \alpha}&=\frac14 e^{iE_0t}\langle c_i e^{-iH_F t} c_i \rangle.
\end{aligned}
\end{equation}

In our approach, the flip of the value of $ ib_i^\alpha b_j^\alpha$ in the Hamiltonian for the time-evolution operator, as seen in the exact case \cite{Baskaran2007}, occurs because we recompute the mean-field parameters for the state on which the Hamiltonian is acting.  In Ref.~\onlinecite{Knolle2018}, the flip occurs due to the anticommutation relations between $b_i^\alpha$ and a newly introduce $\mathbb Z_2$ link variable. Despite agreeing for the exact case, we will see that these two different approaches predict quite different physics in the presence of perturbations. In addition to $S(q,\omega)$, this approach will also reproduce the correct value of the flux gap via computing $\langle b_i^\alpha H b_i^\alpha\rangle$ \cite{Knolle2018}.

\subsection{Finite magnetic field}

One of the advantages of our approach to computing $S(\pmb q,\omega)$ is the ability to treat generic perturbations. In Ref.~\onlinecite{Knolle2018}, it was crucial that the mean-field decoupled Hamiltonian does not mix the $c_i$ and the $b_i^\alpha$'s.  However, a magnetic field is a very natural perturbation, and our approach immediately generalizes.

Firstly, the mean-field decoupled Hamiltonian will now be
\begin{widetext}
\begin{equation}
\begin{aligned}
    H &= H_\text{MF}' = \underbrace{H_c + H_b + H_{bc}}_{H_{bc}'}+ \underbrace{H_C + H_C'}_{H_C''} =\frac18 \sum_{a,b} \psi_a M_{ab} \psi_b + H_C''
\end{aligned}
\end{equation}
\begin{equation}
\begin{aligned}
     H_{bc} &= \frac14 \sum_{\langle i j\rangle_\alpha}\left[ \sum_\beta (J+K \delta_{\alpha,\beta})  \left(ic_i b_j^\beta \langle i c_j b_i^\beta\rangle + ic_j b_i^\beta \langle i c_i b_j^\beta\rangle  - ic_i b_i^\beta \langle i c_j b_j^\beta \rangle-ic_j b_j^\beta \langle i c_i b_i^\beta \rangle  \right) \right. \\
    + & \left.\sum_{\beta \ne \bar \beta \ne \alpha} \Gamma \left(ic_i b_j^\beta \langle i c_j b_i^{\bar \beta}\rangle + ic_j b_i^{\bar \beta} \langle i c_i b_j^{\beta}\rangle - ic_i b_i^{\bar \beta} \langle i c_j b_j^{   \beta} \rangle  - ic_j b_j^{\beta} \langle i c_i b_i^{\bar \beta} \rangle \right) \right] + \frac12 \sum_i \sum_\alpha h_\alpha i c_i b_i^\alpha 
\end{aligned}
\end{equation}
\begin{equation}
\begin{aligned}
    H_C' &= \frac14 \sum_{\langle i j\rangle_\alpha}\left[ \sum_\beta (J+K \delta_{\alpha,\beta})  \left(  \langle ic_i b_i^\beta\rangle  \langle i c_j b_j^\beta \rangle - \langle ic_i b_j^\beta\rangle  \langle i c_j b_i^\beta\rangle \right) + \sum_{\beta \ne \bar \beta \ne \alpha} \Gamma \left( \langle ic_i b_i^\beta\rangle  \langle i c_j b_j^{\bar \beta} \rangle -\langle ic_i b_j^\beta\rangle \langle i c_j b_i^{\bar \beta}\rangle  \right)  \right]
\end{aligned}
\end{equation}
\end{widetext}
where $H_c$, $H_b$, and $H_C$ are defined above. Since all the Majorana's are being intermixed,  we introduced $\psi^T= (c_1,...,c_{2N},b_1^x,...,b_{2N}^x,b_1^y,...,b_{2N}^y, b_1^z,...,b_{2N}^z)$. For ease of notation, we will let $(b^0_i,b^1_i, b^2_i, b^3_i) = (c_i, b^x_i, b^y_i, b^z_i)$ so that $\psi_{i_\alpha} = b_i^\alpha$ where $i_\alpha = i+2N\alpha$.

Secondly, we are going to evolve the state $|\Psi \rangle = c_j b_j^\beta |v\rangle$ in time, and we will need to compute the correlators like
    $i\langle b_i^\alpha c_i \psi_j(t) \psi_k(t) c_i b_i^\alpha \rangle$.
To numerically evaluate this, we just repeatedly apply Wick's theorem in the same way as before. 

Lastly, we need to evaluate the expression
\begin{equation}
\begin{aligned}
    S_{ij}^{\alpha \beta}(t) &= -\frac14 e^{iE_\text{MF}t-i\phi(t)} \langle c_i b_i^\alpha e^{-iH_{bc}'(\mathcal M_t)} c_j b_j^\beta\rangle\\
    e^{-iH_{bc}'(\mathcal M_t)} &= \prod_n e^{-i \Delta t H_{bc}'(t_n)}; \qquad \phi(t) = \int_0^t ds H_C''(s).
\end{aligned}
\end{equation}
In Appendix~\ref{app:Pfaf}, we prove the formula 
\begin{equation}\label{eq:magSij}
\begin{aligned} S_{ij}^{\alpha \beta}(t) &= - \frac14\sqrt{\text{det} X} e^{iE_\text{MF}t -i\phi(t)}\\
&\times \left[ ( UU^\dagger - U F U^T)_{ii_\alpha} (\tilde U \tilde U^\dagger - \tilde U F \tilde U^T)_{jj_\beta }\right. \\
    &-  ( U\tilde U^\dagger - U F \tilde U^T)_{ij} ( U \tilde U^\dagger - U F \tilde U^T)_{i_\alpha j_\beta}
    \\ &\left.+ ( U\tilde U^\dagger - U F \tilde U^T)_{ij_\beta} ( U \tilde U^\dagger - U F \tilde U^T)_{i_\alpha j} \right]
\end{aligned}
\end{equation}
where $\hat U = e^{i M_t} U$ and $X$ and $F=-X^{-1}Y$ are defined from Eq.~\eqref{eq:bararef}. Additionally, in this expression, the multiplication of matrices, $AB$, only involves the first $N$ columns of $A$ and the first $N$ rows of $B$, even if $A$ and $B$ are $2N \times 2N$ matrices.

There is one additional subtlety, however. In a magnetic field, $\langle S_i^\alpha\rangle$ can develop an expectation. Then, $S^{\alpha \beta}(q,\omega) = \tilde S^{\alpha \beta}(q,\omega) + \delta(\omega)\delta(q) \langle S_i^\alpha\rangle \langle S_j^\beta \rangle$. We therefore only really want to calculate
\begin{equation}
    \tilde S_{ij}^{\alpha \beta}(t) = S_{ij}^{\alpha \beta}(t)- \langle S_i^\alpha(t)\rangle \langle S_j^\beta \rangle.
\end{equation}
If we focus on the first term of Eq.~\eqref{eq:magSij}, we see that it can alternatively be written
\begin{equation}
     T_1 = -\frac14 \langle  U^\dagger(0,t) \psi_i \psi_{i_\alpha} \mathcal U(t,0)\rangle \frac{\langle  \mathcal U(t,0) \psi_j \psi_{j_\beta} \rangle}{\langle \mathcal U(t,0) \rangle}.
\end{equation}
Remember, though, that $\mathcal U(t,0) = e^{-iH'_{bc}(\mathcal M_t)-i\phi(t)}\approx U(t,0)$ is just an approximation for the true time-evolution operator.  Using the fact that the ground state should be an eigenstate of $U(t,0)$,  we undo the approximation and find $T_1=\langle S_i^\alpha(t)\rangle \langle S_j^\beta\rangle$. Therefore, $\tilde S_{ij}^{\alpha \beta}(t)$ simply involves the last two terms of Eq.~\eqref{eq:magSij}.

If we do not cancel the term exactly, then when computing $\tilde S^{\alpha\beta}= (q=0,\omega)$ the small approximation on every site gets amplified by the number of sites. A percent-level error then translates to a large discrepancy.

\section{Results}\label{sec:results}

\begin{figure}[tbp!]
    \centering
    \includegraphics[width=0.47\textwidth]{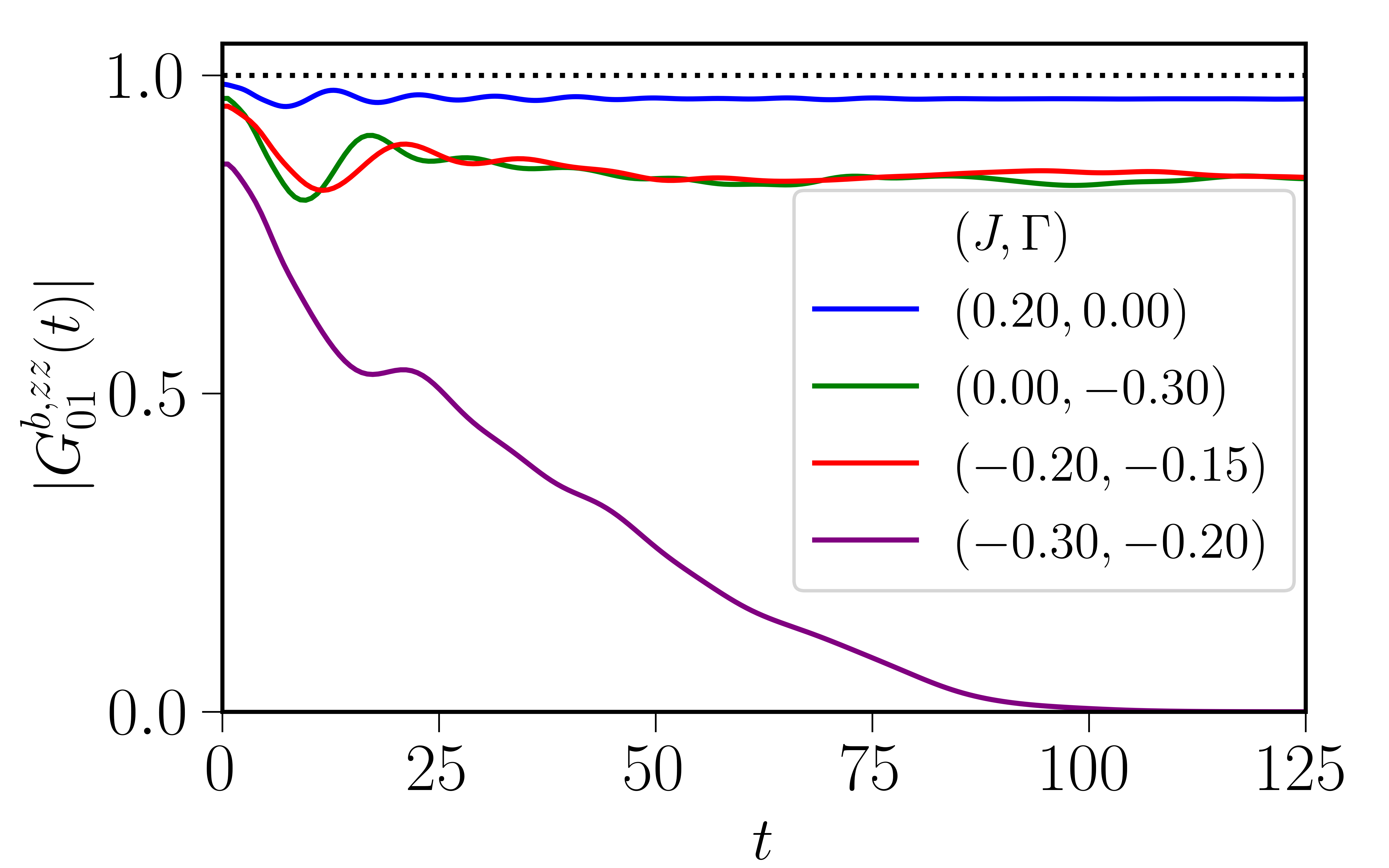}
    \caption{(Color online) We plot $|G^{b,zz}_{01}|$, where sites $0$ and $1$ are connected by a $z$ bond, for a variety of parameters. For small parameters, the asymptotic value as $t\to \infty$ is not substantially different than the starting value. Only when both $J$ and $\Gamma$ are large do we see the value drop, which we can interpret as fluxes become mobile \cite{Knolle2018}. The dashed line indicates the exact ($J=\Gamma=0$) result, and the system sizes are the same as in Fig.~\ref{fig:Sqvt}.}
    \label{fig:Gbvt}
\end{figure}

\begin{figure*}[tbp!]
    \centering
    \includegraphics[width=0.95\textwidth]{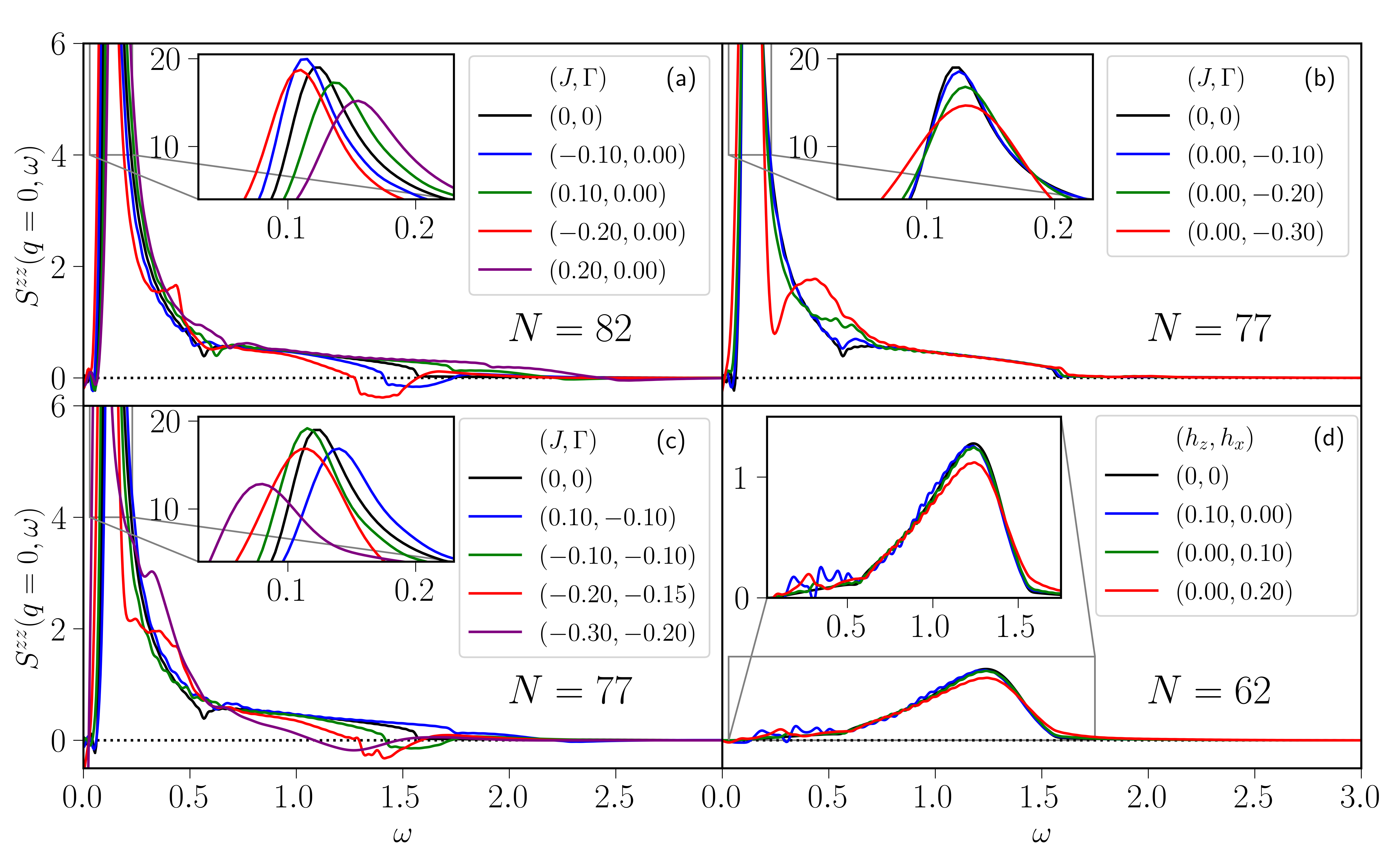}
    \caption{(Color online) We plot $S(q,\omega)$ for a variety of parameters for an $N \times N$ unit cell system. The exact result (black line) is the result from a $N=100$ system and the other parameters have $N$ as specified in each of the panels. In (a), we consider the effect of $J\ne0$ and $\Gamma=0$ and see that the primary effect is shifting the features from the exact case to higher or lower energy. In (b), we plot the same but for $J=0$ and $\Gamma\ne 0$. Beyond some minor adjustments to the peak, the main effect seems to be to smooth out the kink in the exact result. For (c), we see the combination of both $J\ne0$ and $\Gamma\ne 0$ and, for small parameters, the two effects seem roughly additive. For larger parameters, as the flux becomes mobile, there are more substantial changes. In (d), $J=\Gamma=0$ but we consider the effect of a magnetic field in the $z$ direction and $x$ direction.  Due to a smaller time-step, we are not able to consider as large of systems, and so we multiply $S(\pmb q,t)$ by a Gaussian of width $\sigma=60$, equivalent to convolving $S(\pmb q,\omega)$ with a Gaussian of width $1/\sigma$. The main effect of the magnetic field that we see is a smoothing of the high-energy features, and some oscillatory features at low-$\omega$.   We pick $\Delta t$ small enough to ensure convergence (see Appendix~\ref{app:conv}). }
    \label{fig:Sqvt}
\end{figure*}

One limiting factor in the numerics is finite size determined by how long it takes for the Majorana's to travel across the entire system. In the ground state for $\pmb h=0$, the $c$ fermions experience an effective coupling of $\tilde K= (K+J)\langle i b_i^z b_j^z\rangle  + 2J \langle i b_i^x b_j^x\rangle + 2 \Gamma \langle i b_i^x b_j^y\rangle$ giving a speed of $3\tilde K/4$ \cite{Minakawa2020}. A system with $N \times N$ unit cells will then experience finite size effects at roughly $t = 4N/(3\tilde K)$. The only other knob we turn, for a given set of parameters is $\Delta t$, and we ensure that decreasing $\Delta t$ or increasing $N$ has minimal effect on the resulting $S(q,\omega)$ plots. We additionally avoid $N$ that are multiples of $3$ to avoid the gapless points in the Majorana spectrum at the $K$ points \cite{kitaev2006} as they introduce additional complications to the numerics. For additional discussion of convergence, see Appendix~\ref{app:conv}. The finite size effects makes it most difficult to probe small $\omega$, which are also least accessible for inelastic neutron scattering experiments. 

We are primarily interested in computing the results for parameters that we expect to be in the Kitaev phase. For varying $J$ and $\Gamma$, we use the phase diagrams produced via exact-diagonalization on 24 sites in Ref.~\onlinecite{Rau2014Generic}, however we additionally inlcude points at larger $|J|$ when $\Gamma=0$ and vice-versa to highlight the effects that each perturbation has individually. We focus on the ferromagnetic Kitaev model $(K=1)$ as it has larger parameter space when $J,\Gamma \ne 0$, but the qualitative results hold true for $K=-1$. 

One of the main differences between our results and those of Ref.~\onlinecite{Knolle2018} is the flux remains fixed much longer. There are two ways that we can probe this: either by the time evolution of the mean-field paramter $i\langle b_j^\alpha b_i^\alpha(t) b_j^\alpha(t) b_j^\alpha  \rangle$ or by the $b$ component of Eq.~\eqref{eq:SQTDMFT}, $G^{b,\alpha \beta}_{ij}(t) = \langle b_i^\alpha e^{-iH_b^\alpha(\mathcal M_t^b)}b_j^\beta\rangle$. We will use the former as a more direct comparison with Ref.~\onlinecite{Knolle2018}.

We plot $G_b^{zz}(t)$ in Fig.~\ref{fig:Gbvt} and see that even for fairly large perturbations, the flux remains fixed. Only when both $J$ and $\Gamma$ are substantial does the flux begin to move, consistent with the finding of Ref.~\onlinecite{Song2016} \footnote{It is possible very large $J$ or $\Gamma$ individually would be enough to make the fluxes mobile, by this definition, but we have checked for $J=0.4$ and $\Gamma=-0.3$ have not seen this effect}.

We now plot $S(q=0,\omega)$ in Fig.~\ref{fig:Sqvt} for a variety of parameters. In total, we see that the perturbations have only a small effect on the exact result. The Heisenberg term, $J$, primarily moves the features to higher or lower $\omega$, depending on the sign, but the overall qualitative features are the same. For $\Gamma$, there is more power near the kink in the exact result and less power at the peak. When combined, we get some of both features, but, overall, the results are less dramatically different than those found in Ref.~\onlinecite{Knolle2018}.

For the magnetic field, we consider the antiferromagnetic model $K=-1$ as the ferromagnetic model changes phase with $h=0.042$\cite{Nasu2018static} when the magnetic field is aligned with one of the three spin-axes. We additionally find it useful to use a higher order time-evolution scheme \cite{tanaka2010} as the time-step necessary for convergence needs to be smaller.  In the presence of a magnetic field, we can no longer separate the $c$ and $b$ Majorana's, and therefore cannot compute $G_b$.

Due to the smaller time-step, it is difficult to get to as large of system sizes and a well-converged $S(\pmb q,\omega)$, so we multiply $S(\pmb q,t)$ by a Gaussian of width $\sigma=60$. In Fig.~\ref{fig:Sqvt}(d), we plot some results for a magnetic field in the $z$ or $x$ direction. We still find only small effects, such as a smoothing out of high-energy features and oscillatory features at low-$\omega$. In Appendix~\ref{app:compdmrg}, we consider a field in the [111] direction on a cylinder geometry and find similar modifications, like found in Ref.~\onlinecite{Gohlke2018}.

\section{Discussion}\label{sec:disc}

\begin{figure*}[tbp!]
    \centering
    \includegraphics[width=0.90\textwidth]{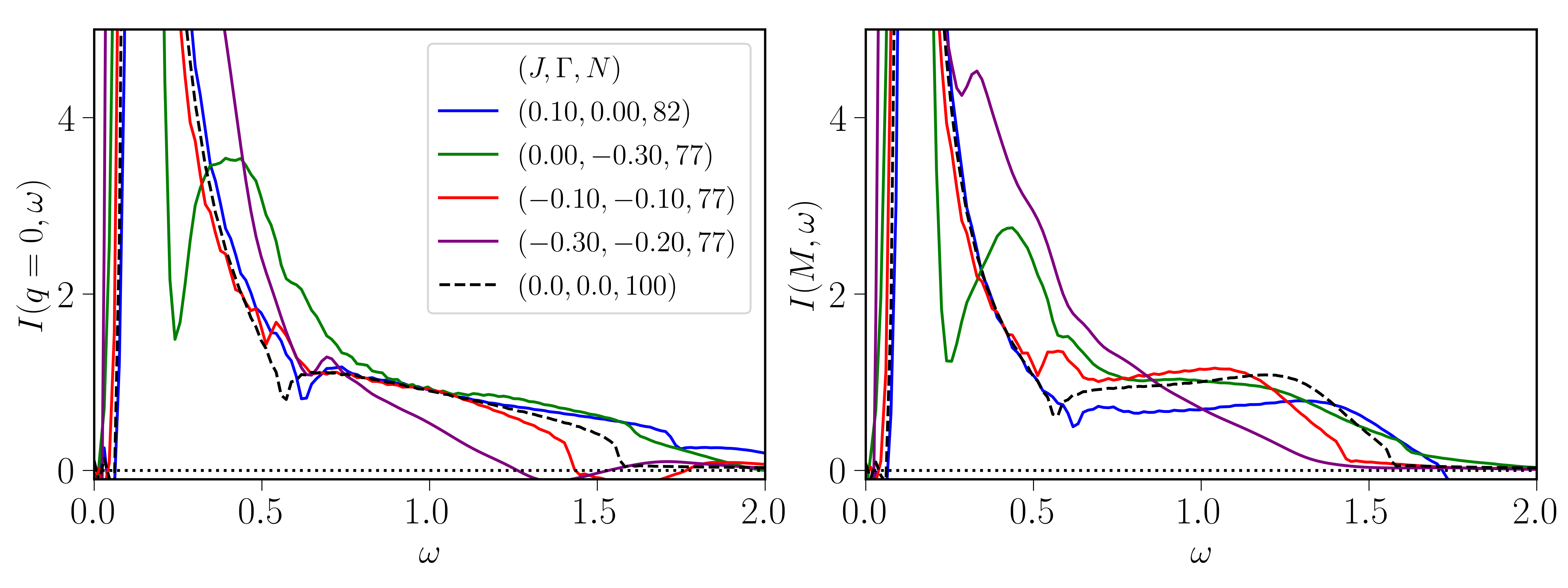}
    \caption{(Color online) We plot the INS intensity (in arbitrary units) at the (a) $\Gamma$ and (b) $M$ point. The legend specifies the size of the system, $N$ used for each parameter set, and when $N$ is not divisible by two, we use the point slightly off of the $M$ point which satisfies the boundary conditions. For smaller perturbations, the features of the exact result are not substantially modified. }
    \label{fig:INS}
\end{figure*}

The most immediate use of our results would be to compare directly with experiments on $\alpha$-RuCl${}_3$ or other Kitaev materials where inelastic neutron scattering (INS) has been performed. We can compute the INS signal with 
\begin{equation}
    I(\pmb q,\omega)\sim f(q)^2 \sum_{\alpha,\beta} \left(\delta_{\alpha\beta}-\frac{q_\alpha q_\beta }{q^2}\right)S^{\alpha\beta}(\pmb q,\omega)
\end{equation}
where we follow Ref.~\onlinecite{laurell2020} in averaging over $q_z$ (assuming that $S^{\alpha\beta}(\pmb q,\omega)$ is independent of $q_z$) as is done in experiment and in approximating the form factor, $f(q)=e^{-q^2 c}$ with $c=(0.25 \times 4\pi)^{-2} \text{ \AA}^2$  to fit the result of Ref.~\onlinecite{do2017}. Since we are envisioning the Jackeli-Khaliulin mechanism \cite{Jackeli2009} for producing a Kitaev material, the $x-$, $y-$, and $z-$ axes for the spins have out-of-plane components, and we account for that when computing $I(\pmb q,\omega)$. We plot the result for a few parameters in Fig.~\ref{fig:INS}.

The large peak in the exact case is not greatly modified by the perturbations, but the smaller higher energy features are. Our results appear quite far from available INS data on $\alpha$-\ch{RuCl3}  \cite{Banerjee2017,banerjee2018} unless the inclusion of an in-plane magnetic field leads to substantial changes. Furthermore, our approach is only valid in the Kitaev phase, and we therefore hesitate to  compute $S(q,\omega)$ with some of the best candidate spin-Hamiltonians of $\alpha$-\ch{RuCl3} since numerical studies of these models do not support the conclusion that the field-induced spin-liquid is a Kitaev spin liquid \cite{Winter2018,gordon2019}. In identifying and studying other Kitaev materials, the main result of our approach is that the INS signal should be well-captured by the exact Kitaev model.

One major technicality that we have not addressed is the role of gauge. Due to the enlargement of the Hilbert space via the introduction of four Majorana's per spin, we must project the unphysical degrees of freedom away with the operator $P$. The true ground state of the system would then be $P|v\rangle$, and we explore the effect of this in Appendix~\ref{app:gauge}. In total, our approach is consistent with other mean-field treatments in the literature, but more consideration is likely warranted in the future.

One shortcoming of our approximation is that it does not agree with exact bounds. Using the Lehmann representation, it is clear that $S^{zz}(q,\omega)\ge 0$ \cite{Zhang2021_VMC}, and we expect sum rules to be obeyed such as 
\begin{equation}\label{eq:sumrule}
    S^{zz}(\pmb q,t=0) = \frac1{2\pi} \int d\omega S^{zz}(\pmb q,\omega).
\end{equation}
In the former case, we can quantify the disparity by computing $P_\text{neg}=\int d\omega S^{zz}(q,\omega) / (\int d\omega |S^{zz}(q,\omega)|)$, and in the latter case, we compute $P_\text{diff}$, the percent difference between the two sides of Eq.~\eqref{eq:sumrule}. 

In total we get the results plotted in Table.~\ref{tab:errs}. Focusing solely on $q=0$, except for the largest parameter point, we see the error is $\lesssim 5\%$. At the $M$ point, there are larger errors for when $J$ is the only perturbation,  but, otherwise, the same is true. Of course, these discrepancies must go to zero for small perturbations since it must be zero in the exact case.

\begin{table}[]\label{tab:errs}
\begin{tabular}{c | c| c}
$(J,\Gamma,h_z,N)$    & $S(q=0,\omega)$ \% err.         & $S(M,\omega)$ \% err.     \\
\hline
\hline
$(-0.1,0.0,0.0,82)$   & $(1.7,0.5)$   & $(7.5,-16)$    \\
$(0.1,0.0,0.0,82)$    & $(0.42,0.29)$ & $(6.2,-15)$    \\
$(-0.2,0.0,0.0,82)$   & $(3.2,3.0)$   & $(7.5, -14)$   \\
$(0.2,0.0,0.0,82)$    & $(0.65,1.1)$  & $(5.3,-14)$    \\
$(0.0,-0.1,0.0,77)$   & $(0.14,0.46)$ & $(0.12,0.39)$  \\
$(0.0,-0.2,0.0,77)$   & $(0.03,2.0)$  & $(0.02,1.7)$   \\
$(0.0,-0.3,0.0,77)$   & $(0.047,5.1)$ & $(0.035,4.5)$  \\
$(0.1,-0.1,0.0,77)$   & $(0.35,0.66)$ & $(0.10,0.55)$  \\
$(-0.1,-0.1,0.0,77)$  & $(1.4,1.2)$   & $(0.048,0.92)$ \\
$(-0.2,-0.15,0.0,77)$ & $(2.6,4.8)$   & $(0.27,3.6)$   \\
$(-0.3,-0.2,0.0,77)$  & $(4.1,11.4)$  & $(2.0,8.0)$    \\
\end{tabular}
\caption{We list the disagreement with exact bounds for the parameters in the left hand column for $S(q,\omega)$ for $q=0$ and for $q$ at the $M$ point. The error is reported as $(P_\text{neg}, P_\text{diff})$ where $P_\text{neg}$ is the percent of the support that is negative and $P_\text{diff}$ is the percent difference between the left- and right-hand side of Eq.~\eqref{eq:sumrule}. Although some of the errors are in the 10's of percent, most are 5\% or less.}
\end{table}

Additionally, although TDMFT clearly is an important starting point for the Kitaev case,  the computation of the mean-field parameters at each time step and the exponentiation of $\mathcal M_t$ greatly increases the cost of computing dynamical quantities. For other systems, this may make TDMFT impracticable. When, then is it necessary to apply TDMFT instead of evolving in time under the ground-state mean-field decoupled Hamiltonian? We leave a detailed analysis for future work, but the importance for the Kiteav model seems to be connected to the localization of the fluxes. In fact, if one were to apply TDMFT not to $S_{ij}^{\alpha \beta}(t)$ but instead directly to $S^{\alpha\beta}(\pmb q,t)=\langle S_{-\pmb q}^\alpha(t) S_{\pmb q}^\beta\rangle$ where $S_{\pmb q}^\alpha(t) = \sum_i S_i^\alpha(t) e^{-i\pmb x_i \cdot \pmb q}$, the exact result would not be recovered. Indeed, in the latter case the single flux being flipped would be distributed across the lattice and the mean-field value of $i\langle S_{\pmb q}^\alpha b_i(t) b_j(t)S_{-\pmb q}^\alpha \rangle/\langle S_{\pmb q}^\alpha S_{-\pmb q}^\alpha\rangle $ would be uniform and unaffected in the thermodynamic limit. Although this introduces ambiguity into how TDMFT should be applied, it is clear that, for the Kitaev model, the former is the correct starting point. If the latter can be argued to be the better starting point, then TDMFT will produce the same results as time evolution under the mean-field ground state in the thermodynamic limit.

\section{Conclusions}\label{sec:conc}

In this paper, we have rigorously developed time-dependent Majorana mean-field theory, as introduced by \cite{Nasu2019,Taguchi2021,Minakawa2020} and applied the technique directly to compute dynamic correlators. This approach immediately reproduces the exact results of the Kitaev model, and we therefore expect it to qualitatively capture the effects of perturbations. Although we have only considered the Kitaev model here, our approach applies generally to any mean-field decoupled (or quadratic) Majorana system, and it should be generalizable to any mean-field decoupled fermionic or bosonic system. 

In comparing and contrasting our approach with Ref.~\onlinecite{Knolle2018}, we both recover the exact result in the absence of perturbations, but our approach immediately extends to the case with perturbations without any additional approximations. Furthermore, the $\mathbb Z_2$ link variable that Ref.~\onlinecite{Knolle2018} introduces provides feedback between the $b_i^\alpha$ and the $c_i$ Hamiltonian, but our approach naturally includes both that and the feedback between the $c_i$ and $b_i^\alpha$ Hamiltonian. Additionally, since we treat $c_i$ and $b_i^\alpha$ on the same footing, we can accommodate any perturbation, and we are able to recover an explicit expression for their $V^\alpha_{A0}(t)=(K+J)(i\langle b_j^\alpha b_i^\alpha(t) b_j^\alpha(t) b_j^\alpha  \rangle-i\langle b_i^\alpha b_j^\alpha\rangle)/4$, which they approximated via a Heaviside step function. With the inclusion of fewer approximations, our results indicate that the features of the exact model are not significantly modified in the presence of small perturbations, in contrast to previous results \cite{Song2016}. 

We also emphasize that our approach will agree with exact results of the Kitaev model for \textit{any} dynamic correlator. In the exact case, the correlators will be evaluated by commuting any $\prod b_i^\alpha$ to the left or right to act on the ground state $|v\rangle$, which is equivalent to recomputing the mean-field parameters for the state $\prod b_i^\alpha |v\rangle$. One natural future direction then would be to apply our approach to the current-current correlator necessary to compute $\kappa_{xx}$ and $\kappa_{xy}$ \cite{Pidatella2019,Nasu2017}. 

\section{Acknowledgements} We thank James Analytis, Shubhayu Chatterjee, Tomohiro Soejima, and Vikram Nagarajan for fruitful discussions.  T.C. was supported by NSF DMR-1918065 and an NSF Graduate Fellowship under Grant No. DGE 2146752.  J.E.M. was supported by the Quantum Science Center (QSC), a National Quantum Information Science Research Center of the U.S. Department of Energy (DOE).  J.E.M. acknowledges additional support by a Simons Investigatorship.
\appendix

\section{Time-evolution in Majorana mean-field theory: a comparison}\label{app:tdmftvother}

When applying mean-field theory to time evolution of states $|\Psi\rangle$, one starting point is to use
\begin{equation}
    U(t,0)|\Psi\rangle = e^{-iHt}|\Psi \rangle = e^{-iH_\text{MF,GS}t}|\Psi\rangle
\end{equation}
where $H$ is some arbitrary Hamiltonian and $H_\text{MF,GS}$ is the mean-field decoupled Hamiltonian where the mean-field parameters are determined in the ground state. For states near the ground state, this approximation might be reasonable.

We can compare this kind of evolution to TDMFT by doing the following. First, we follow Ref.~\onlinecite{Nasu2019} by computing $S(q=0,\omega)$ via a quantum quench from a small magnetic field. In this case, we write the Jordan-Wigner transformed Kitaev Hamiltonian in an out-of-plane magnetic field as
\begin{equation}
\begin{aligned}
    H(h) &= -i\frac{K}{4} \sum_{j\in A}\sum_{ \alpha=x,y}a_jb_{j+\alpha}
    \\
    &-\frac{K}{4}ia_jb_{j+\hat z}i\bar a_j \bar b_{j+\hat z}-i\frac{h}2 (a_j \bar a_j - b_{j+z} \bar b_{j+z})
\end{aligned}
\end{equation}
In this rewriting, the conserved quantity at $h=0$ is $i\bar a_j \bar b_{j+\hat z} = \pm 1 = \bar \Phi_{j}$, and the ground state has $\bar \Phi_j = \bar \Phi = 1$. 

Now, we find the ground state of $H(h)$ for small $h$, and compute the time evolution of the ground state under the Hamiltonian $H(h=0)$.  By computing $\langle M^z(t)\rangle = \langle S^z(t))_i\rangle$, we can compute \cite{Nasu2019}
\begin{equation}\label{eq:Sqquench}
    S(q=0,\omega) = \frac{1}{N}\sum_{i,j}\int dt e^{i\omega t} \langle S_i(t)S_j\rangle = \lim_{h\to 0} 2\frac{\tilde M^z(\omega)}{h}
\end{equation}
where $\tilde M^z(\omega)/h=\omega \text{Re}\left[\int_0^\infty dt e^{i(\omega+i\eta)t}M^z(t)\right]$ with $\eta\ll 1$. 

We can evolve in time in two ways--the first is TDMFT \cite{Nasu2019} and the second is to instead evolve with $H_{MF,GS}$, which in this case is $H=H_0 = -i(K/4) \sum_{j\in A}\sum_{\alpha} a_j b_{j+\alpha}$. In the former case, we self-consistently compute the expectations $A = \langle i a_j \bar a_j\rangle$, $B = \langle i b_j \bar b_j\rangle$, $\langle ia_j b_{j+\hat z}\rangle$, $\langle i\bar a_j \bar b_{j+\hat z}\rangle$, $\langle i b_j \bar a_j\rangle$, and $\langle i a_j \bar b_j\rangle$. 

To do the numerics, we Fourier transform and perform time-dependent mean-field theory in $k$-space. Since each $(k,-k)$ pair is independent, we just need to keep track of the $4\times 4$ matrix that provides the time-evolution operator for that pair. To compute the $k$-integrals for expectation values, we keep track of $N_k^2$ points in the Brillouin zone that are distributed as per Gaussian quadrature, and we take $N_k$ as large as the numerics will allow. 

We evolve for a time $t|K|=2.5 \times 10^4$ using the Euler step method \cite{tanaka2010}, $\eta/|K| = 7.5\times 10^{-4}$, and our initial magnetic field is $h/K=0.0015$. Additionally, we average $S(q=0,\omega)$ over windows of $\Delta \omega = 0.01 K$ because of rapid oscillations. We are able to essentially reproduce the TDMFT curve from Ref.~\onlinecite{Nasu2019} and we derive an analytic result below that matches evolution under $H_\text{MF,GS}$. 

\begin{figure}[tbhp!]
    \centering
    \includegraphics[width=0.45\textwidth]{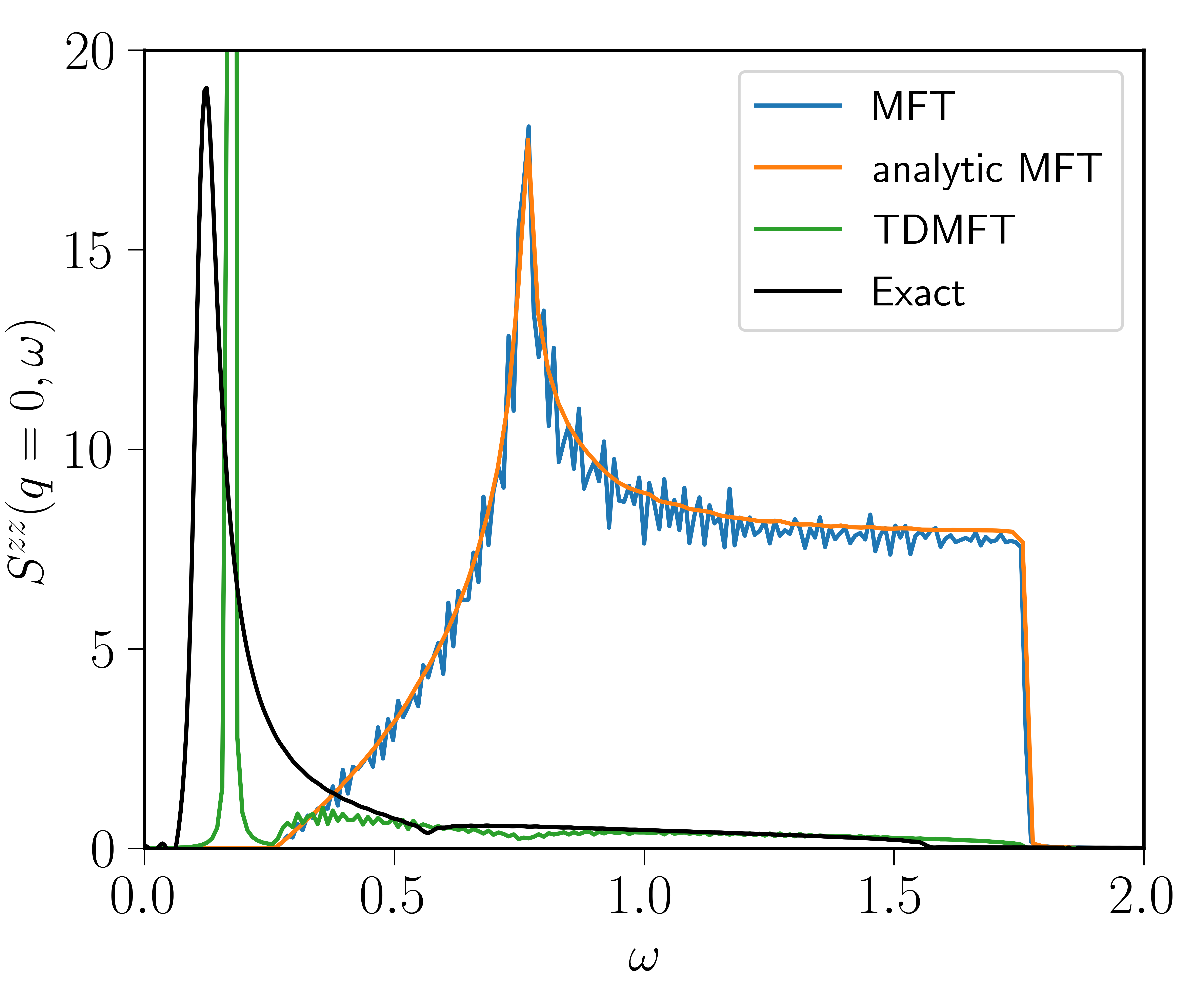}
    \caption{(Color online) We compare computing $S^{zz}(q=0,\omega)$ with a quantum quench using TDMFT and evolving under $H_\text{MF,GS}$, which we refer to as MFT. When compared to the exact answer (black curve), it is clear that TDMFT does substantially better. Using the augmented mean-field theory of Ref.~\onlinecite{Knolle2018} produces the same curve as MFT. We include the analytic result Eq.~\eqref{eq:Sqquench_an}, which demonstrates the numerics work. Here, we use $N_k=200$ and the rest of the parameters are given in the text, and the exact result is evaluated using the Pfaffian method of Ref.~\onlinecite{Knolle2015} for 100x100 unit cells. }
    \label{fig:TDMFT_MFTcomp}
\end{figure}

We see in Fig.~\ref{fig:TDMFT_MFTcomp} that TDMFT is able to capture all the qualitative features of the exact result whereas evolution under $H_\text{MF,GS}$, labeled as MFT, produces a completely different result. This plot heavily implies that the starting point of understanding time-evolution in mean-field theory should be TDMMFT.

\subsection{Analytic MFT result}

In addition to numerics, we can exactly compute $M^z(t)$ in the case that we are evolving under $H_\text{MF,GS}=H_0$. 
First, we observe that the state we are evolving is the ground state of $H(h)$. In the limit $h\to 0$, we write the self-consistent value of $\tilde h = h + KA/2$. We can write the Hamiltonian as $H(h) = H_0 - \tilde h N M^z$ (where $N$ is the number of sites), and treat the second term as a perturbation. The ground state can be written as
\begin{equation}
    |\Psi\rangle = |0\rangle - \tilde h N \sum_{n\ne 0} |n\rangle \frac{\langle 0 | M^z |n \rangle }{E_0-E_n}.
\end{equation}
Our next task is to determine  which states $|n \rangle$ have non-zero values of $\langle 0 | M^z | n\rangle$. By going to Fourier space, the resulting Hamiltonian is given by
\begin{equation}
\begin{aligned}
    H = \frac12 \sum_k &\begin{pmatrix} a_{-k} & b_{-k} \end{pmatrix} \begin{pmatrix} 0 & S_k \\
    S_k^* & 0 \end{pmatrix} \begin{pmatrix} a_k \\ b_k \end{pmatrix} 
    \\
    + &\begin{pmatrix} \bar a_{-k} & \bar b_{-k} \end{pmatrix} \begin{pmatrix} 0 & T_k \\
    T_k^* & 0 \end{pmatrix} \begin{pmatrix} \bar a_k \\ \bar b_k \end{pmatrix},
\end{aligned}
\end{equation}
where $a_i = \sqrt{\frac{2}{N}}\sum_k e^{ikr_i} a_k$ and $a_k^\dagger = a_{-k}$. Here, $S_k = -iJ (e^{-ik\cdot n_x} + e^{-ik\cdot n_y} + 1)e^{i \delta_k}/2$ and $T_k = -i J \Phi e^{i\delta_k}/2$ where $n_{x/y} = (\pm 1/2, \sqrt{3}/2)$, $\delta_k = k_y/\sqrt{3}$, and $\Phi = i\langle a_i b_{i+z}\rangle \approx -0.5249$. We now diagonalize these Hamiltonians to get $H=\sum_k |S_k| (f_k^\dagger f_k-1/2) + |T_k| (\bar f_k^\dagger \bar f_k-1/2)$. Rewriting $M^z$ in the $f_k$ basis, and acting on the state vacuum in that basis, we get
\begin{equation}
\begin{aligned}
    M^z|0\rangle &= \frac{1}{N}\sum_i ia_i\bar a_i-i b_{i+z}\bar b_{i+z} |0\rangle  
    \\
    &= \frac{1}{2N} \sum_k -\left( \frac{|S_k||T_k|}{S_k^* T_k} +1\right) f_{-k}^\dagger \bar f_k^\dagger |0\rangle. 
\end{aligned}
\end{equation}
Therefore, the $N/2$ states we need to consider are $|k\rangle = f_{-k}^\dagger \bar f_k^\dagger |0\rangle$ (one for each $k\in 1BZ$), and the energy is $E_k = |S_k| + |T_k|+E_0$ 

Now, it is a straightforward computation that
\begin{equation}
\begin{aligned}
    \frac{\langle M^z(t)\rangle}{h} &= N \frac{\tilde h}{h}\sum_{k \in 1BZ} \frac{\langle 0 |M^z(t)|k\rangle \langle k | M^z(0)| 0 \rangle + \text{H.c}}{E_k-E_0}
    \\
    &=N\frac{\tilde h}{h}\sum_{k\in 1BZ} \frac{|\langle k | M^z(0)| 0 \rangle|^2}{E_k-E_0}(e^{i(E_0-E_n)t} + \text{H.c})
\end{aligned}
\end{equation}
which after integration (and taking only $\omega>0$)
\begin{equation}\label{eq:Sqquench_an}
\begin{aligned}
  &  S(q=0,\omega) = 2 N \frac{\tilde h}{h} \sum_{k \in 1BZ}\langle k |M^z(0)|0\rangle^2\delta(\omega-E_n-E_0) 
    \\
    &= \frac{\tilde h}{h}\frac{1}{N} \sum_{k \in 1BZ}\left(1+\text{Re}\left[\frac{|S_k||T_k|}{S_k^* T_k}\right]\right)\delta(\omega - |S_k|-|\Phi|/2).
\end{aligned}
\end{equation}

The final term we evaluate by rewriting $\delta(x) = \eta/(x^2+\eta^2)$ where $\eta$ plays the same role as in Eq.~\eqref{eq:Sqquench}. This expression resembles the density of states, but has some additional energy dependence. As seen in Fig.~\ref{fig:TDMFT_MFTcomp}, the exact result and numerics are in good agreement.

\section{Comparison with DMRG} \label{app:compdmrg}

In this appendix we will compare our results using TDMFT with the density-matrix renormalization group (DMRG)\cite{White1992}. DMRG results are ``exact'' if the bond-dimension $\chi$, the size of the matrices, goes to infinity. See e.g. \cite{schollwock2011} for a review of the technique.

\begin{figure*}[tbp!]
    \centering
    \includegraphics[width=0.47\textwidth]{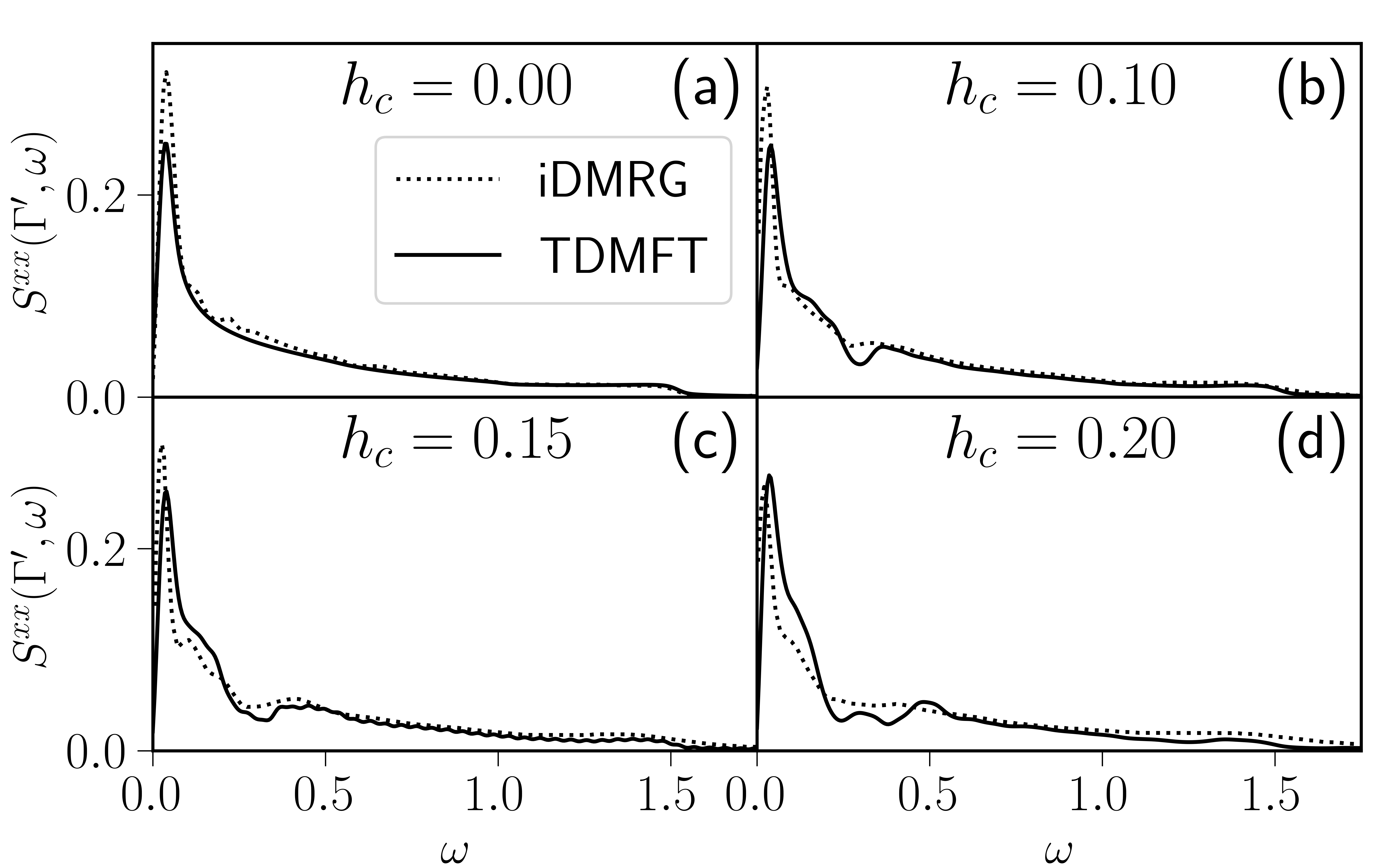}\includegraphics[width=0.47\textwidth]{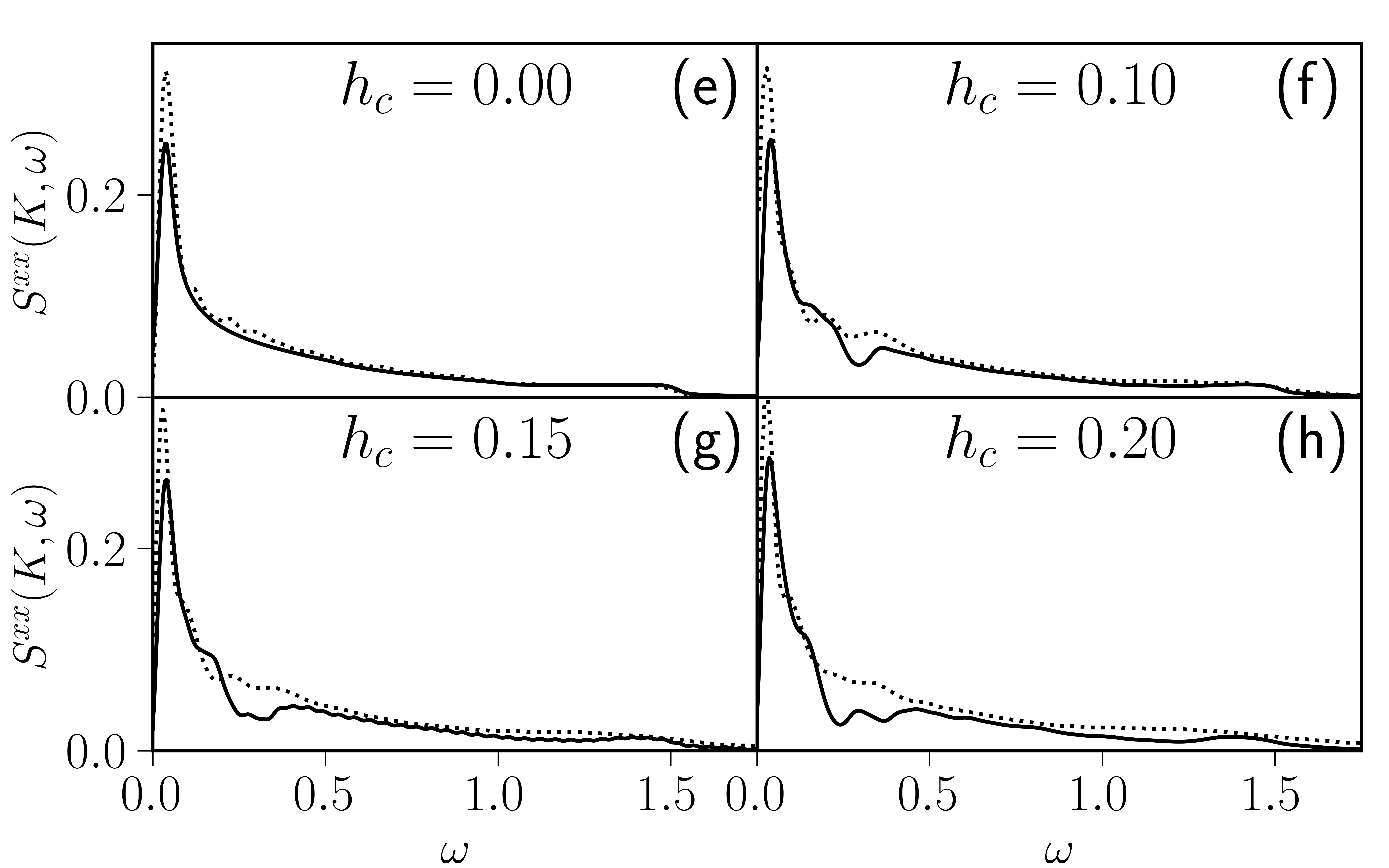}
    \caption{We plot the results of TDMFT vs. iDMRG from Ref.~\onlinecite{Gohlke2018} for the $N_y=3$ cylinder. In (a)-(d) [(e)-(h)], we plot $S(\pmb q,\omega)$ at the $\Gamma'$ ($K$) point as definied in \cite{Gohlke2018} for various magnetic field strengths in the [111] direction. In (b) and (c) we see that TDMFT does captures the dominant effects of the magnetic field, but the comparison for (f) and (g) is worse. For (a) and (e), the two curves should agree, and the difference is likely due to error in the iDMRG calculation incurred from having a finite bond-dimension, which may explain some of the discrepancy in plots (f) and (g). Overall, TDMFT seems to be providing qualitatively accurate results. }
    \label{fig:iDMRG_TDMFT_comp}
\end{figure*}

The authors of Refs.~\onlinecite{Gohlke2017} and \onlinecite{Gohlke2018} have applied infinite DMRG to compute $S(\pmb q,\omega)$ in the presence of Heisenberg terms or a magnetic field in the [111] direction. In the latter case, we can directly compare TDMFT to their results in Fig.~6(g) of Ref.~\onlinecite{Gohlke2018}. 

In Fig.~\ref{fig:iDMRG_TDMFT_comp}, we compare the results for $\pmb q=\Gamma'$ and $\pmb q=K$, as defined in their work, and omit $\pmb q=M$ since it is similar to $\pmb q=K$. For TDMFT, we considered a system size of $(N_y,N_x)=(3,152)$ with step size $\Delta t=0.32$, large enough to have negligible finite-size effects, and we multiply $S(\pmb q,t)$ by a Gaussian of width $\sigma=55.8$ as in Ref.~\onlinecite{Gohlke2018}. Our results for $h_c=0$ are really for $h_c=0.003$, but we have checked that this does not effect our forthcoming analysis. We scale our results by an $h_c$-independent constant to match the results of Ref.~\onlinecite{Gohlke2018} at large $\omega$ and $h_c=0$ to account for their normalization of $S(\pmb q,\omega)$. 

Even in the exact case, where the two should methods should agree, there are discrepancies at small $\omega$. These differences are likely due to the finite bond-dimension in the DMRG simulations since larger and larger bond-dimesions are needed to capture longer and longer time behavior \cite{schollwock2011}, as can be seen in the insets of Fig.~3 in Ref.~\onlinecite{Gohlke2017}. 

In light of this, comparing the $h_c\ne 0$ results is not straightforward since the largest discrepancies appear at low $\omega$ where the $h_c=0$ results disagree. Nevertheless, there is reasonable qualitative agreement between the results--at large $\omega$, the features are smoothed out with increasing $h_c$, and similar oscillating features are added at small $\omega$. Additionally, the perturbation $h_c$ only slightly modifies the overall features of $S(\pmb q,\omega)$, consistent with our results.

To include an additional test, we compare our approach and that of Ref.~\onlinecite{Knolle2018} to short-time DMRG evolution. We consider a $2\times N_y \times N_x$ system with periodic boundary conditions in the $N_y$ direction and open boundary conditions in the $N_x$ direction. We time-evolve the system for short times and check convergence in $\Delta t$ and the bond-dimension $\chi$. We use the \texttt{TeNPy} \cite{tenpy} package, and time-evolution is performed by constructing an MPO representation of the time-evolution operator \cite{Zaletel2015}. We are able to get exact agreement in the unperturbed model. The z bond is chosen to be either of the two bonds more closely aligned with the short axis of the cylinder. 

\begin{figure}[tbp!]
    \centering
    \includegraphics[width=0.45\textwidth]{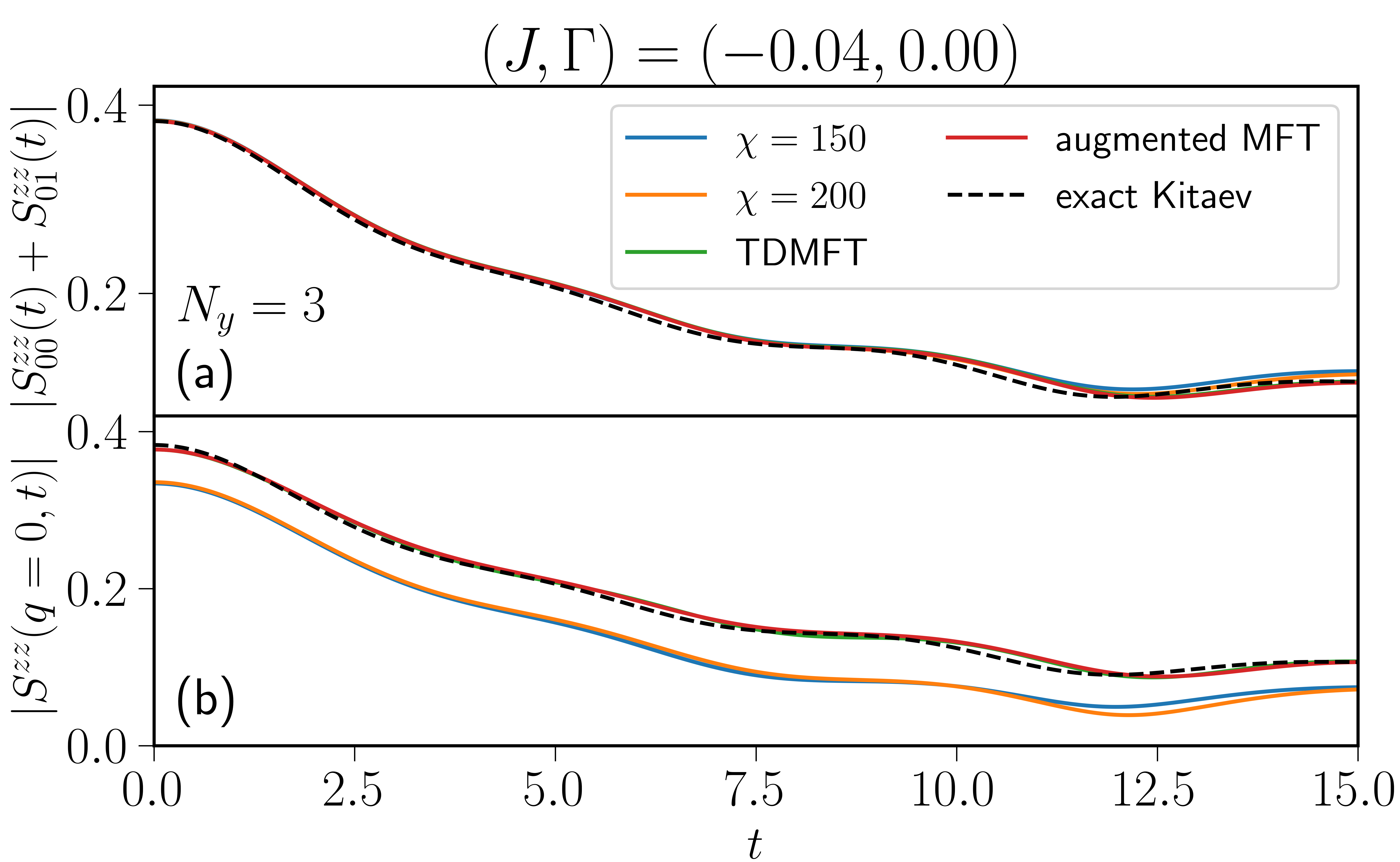}
    \caption{(Color online) In (a) we plot $|S^{zz}_{00}(t)+S^{zz}_{01}(t)|$ vs. t (where site 0 is far from the boundaries of the cylinder and connected to site 1 by a $z$ bond) using DMRG at bond dimension $\chi$, using the augmented MFT of Ref.~\onlinecite{Knolle2018}, and using TDMFT. For reference, we include the exact result from the unperturbed Kitaev point. We see that the magnitude of the sum of the two most important correlators for the unperturbed model are accurately shifted (though we note that the phases disagree).  The DMRG and MFT results only begin to diverge when the DMRG result is no longer converged in bond dimension at around $t\sim 11$. (b)  We plot $S^{zz}(q=0,t) = \sum_i S^{zz}_{0i}(t)$ vs. t computed through the various methods as in (a). There is a large quantitative shift, but the qualitative features agree between the three methods. The shift decreases with increasing cylinder size as seen in Fig.~\ref{fig:DMRGcomp2}.  Here $N_x=20$. }
    \label{fig:DMRGcomp}
\end{figure}

\begin{figure}[tbp!]
    \centering
    \includegraphics[width=0.45\textwidth]{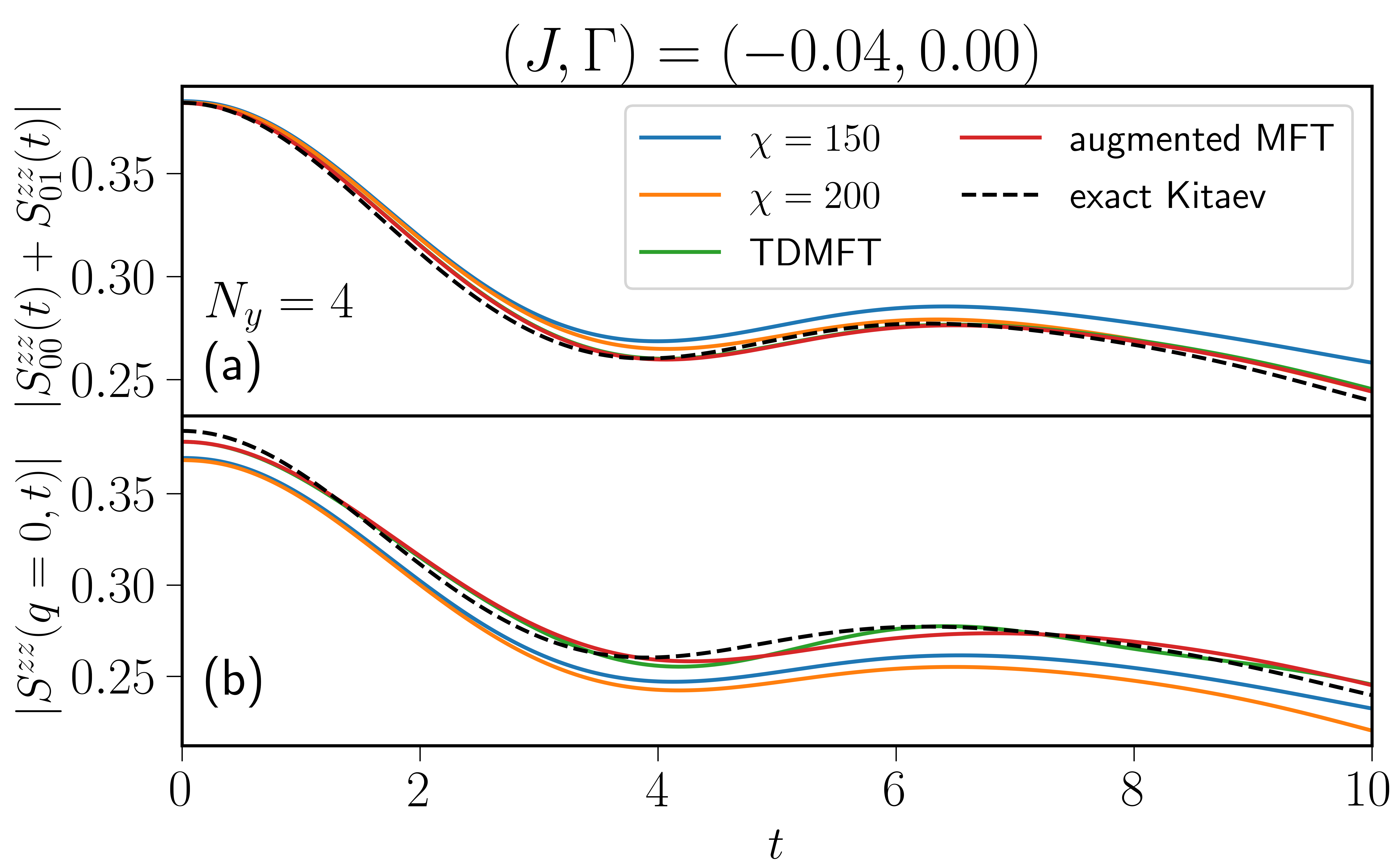}
    \caption{(Color online) We make the same plot as in Fig.~\ref{fig:DMRGcomp} for the $N_y=4$ cylinder. Both results are not well converged in bond dimension, but we notice that $S^{zz}(q=0,t)$ is more quantitatively similar than the $N_y=3$ cylinder implying some of the discrepancy is due to the small circumferences. Here $N_x=16$. }
    \label{fig:DMRGcomp2}
\end{figure}

We 
consider the small perturbation $J=-0.04$, 
and plot the result for two cylinder sizes in Fig.~\ref{fig:DMRGcomp} 
and Fig.~\ref{fig:DMRGcomp2}. 
We plot both $S^{zz}(q=0,t) = \sum_i S_{i0}^{zz}(t)$ and $S^{zz}_{00}(t) + S^{zz}_{01}(t)$ 
where the site 0 is picked to be far from the open boundary conditions and is connected to site 1 by a $z$ bond.
 
For the $N_y = 3$ cylinder in Fig.~\ref{fig:DMRGcomp}, we are able to get to large enough bond dimension to have $t \lesssim 11$ converged. Remarkably, we see that both MFT approaches accurately captures the shift in $|S^{zz}_{00}(t) + S^{zz}_{01}(t)|$, the two correlators that contribute the most in the unperturbed model. However, the phase is not accurately captured (not shown), and when we sum over all sites for $S^{zz}(q=0,t)$, the MFT and DMRG approaches disagree quantitatively but have similar features. The latter point is expected since the overall features must closely match the unperturbed result. For the $N_y=4$ cylinder in Fig.~\ref{fig:DMRGcomp2}, the convergence in bond-dimension is worse, but the quantitative discrepancy between $S^{zz}(q=0,t)$ decreases implying that it is in part due to small cylinder circumferences. 

Taken together, TDMFT compares favorably with iDMRG and DMRG. We were unable to get to large enough bond dimension to directly determine whether TDMFT or augmented MFT is more accurate, but TDMFT extends to the finite field case and the results of Appendix~\ref{app:tdmftvother} show that TDMFT is more broadly applicable.

\section{Evaluating correlators}\label{app:Pfaf}

In order to evaluate Eq.~\eqref{eq:SQTDMFT}, we need to evaluate expressions of the form
\begin{equation}
   I_{ij} =  \langle a_i e^{-iH(\mathcal M)}a_j^\dagger \rangle
\end{equation}
with regards to the vacuum $|v\rangle$ of the operators $\bar a = \frac{1}{\sqrt{2}}U_0^\dagger \vec c_i$.
Our first step is finding the basis $\bar b =S^\dagger \vec {\tilde c}=S^\dagger U_0 \bar a$ such that $\mathcal M = S DS^\dagger$. In that case,
\begin{equation}
\begin{aligned}
    I_{ij} &= \langle a_i (U_0^\dagger S e^{i \mathcal D}S^\dagger U_0)_{j+N,k} \bar a_k e^{-iH(\mathcal M)}\rangle
    \\&=\mathcal T_{j+N,k}\sqrt{\text{det} X}\langle a_i \bar a_k e^{\frac12  a^\dagger_\alpha F_{\alpha \beta} a^\dagger_\beta  }\rangle
\end{aligned}
\end{equation}
here we have used Eq.~\eqref{eq:vac} and $\mathcal T=U_0^\dagger e^{i\mathcal M}U_0$ is the change of basis matrix between $\bar a$ and $\bar a^{(t)}$. There is also an implicit sum over repeated Greek letters (e.g. $\alpha$ and $\beta$) from $1$ to $N$ and repeated Roman letters (e.g. $k$) from $1$ to $2N$. Now, we expand the exponential since only the first two terms will produce non-zero overlaps. We find
\begin{equation}
\begin{aligned}
    I_{ij} &= \mathcal T_{j+N,k}\sqrt{\text{det} X}\left[\delta_{i,k-N} +  \frac12 F_{\alpha \beta}(\delta_{k\alpha}\delta_{i\beta} - \delta_{k\beta}\delta_{i\alpha})\right]\\
    &=\sqrt{\text{det} X}  \left[  X^\dagger_{i,j}+ (X^{-1} Y)_{i,\alpha}Y^*_{j,\alpha}\right]=\sqrt{\text{det} X} X^{-1}_{ij}
\end{aligned}
\end{equation}
where $X$ and $Y$ are related to the four submatrices of $\mathcal T$ as in Eq.~\eqref{eq:vac}. The last step follows because $\mathcal T$ is unitary so $\mathcal T \mathcal T^\dagger = 1 \implies 1= X X^\dagger +Y Y^\dagger = X (X^\dagger +X^{-1} Y Y^\dagger)$. We have thus arrived at Eq.~27 of Ref.~\onlinecite{Knolle2015} without needing to manipulate Pfaffians. 

As noted in the main text, we want to extract the continuous function $\phi (t) = \text{arg} [\text{det}(X)]$, which becomes a very rapidly changing function as system sizes become larger. Fortunately, a large portion of the change in $\phi(t)$ may be canceled from the prefactors $e^{iE_{MF}t}$ and $e^{-i\psi(t)}$ in Eq.~\eqref{eq:SQTDMFT}. 

In the presence of a magnetic field, we need to essentially evaluate
\begin{equation}
\begin{aligned}
    J_{ijkl} &= \langle c_i c_j e^{-iH(\mathcal M)} c_k c_l\rangle 
    \\&= \sqrt{\text{det}X}\langle c_i c_j c_k(-t) c_l(-t) e^{\frac 12 a^\dagger_\alpha F_{\alpha \beta} a^\dagger_\beta}\rangle
\end{aligned}
\end{equation}
with an implicit sum over $p$ and $q$ as before and $c_k(t) = e^{iH(\mathcal M)} c_k e^{-iH(\mathcal M)}$. We introduce the two matrices $ \vec c = \sqrt{2} U \bar a$ and $ \vec c(-t) = \sqrt{2}\hat U \bar a$ and $\hat  U  = e^{i \mathcal M} U$.

By making the following two observations
\begin{equation}
\begin{aligned}
    \langle c_i c_j(-t)\rangle &=  2U_{i\alpha}\hat U_{j,\beta+N}  \langle a_\alpha a_\beta^\dagger \rangle =2 U_{i\alpha}\hat U_{\alpha j}^\dagger  \\
    \langle c_i a_j^\dagger \rangle &= \sqrt{2} U_{i\alpha}\langle a_\alpha a_j^\dagger \rangle = \sqrt{2} U_{ij}
\end{aligned}
\end{equation}
where greek letters are implicitly summed only from $1$ to $N$.
we can easly compute that $J_{ijkl}= \sqrt{\text det X} (J_1 + J_2+J_3)$ where
\begin{widetext}
\begin{equation}
    J_1 =\langle c_i c_j c_k(-t) c_l(-t)\rangle = 4( (UU^\dagger)_{ij} (\hat U \hat U^\dagger)_{kl} -(U\hat U^\dagger)_{ik} (U \hat U^\dagger)_{jl}  + (U\hat U^\dagger)_{il} (U \hat U^\dagger)_{jk} 
\end{equation}
\begin{equation}
\begin{aligned}
    J_2  =\frac12 F_{\alpha \beta}\langle c_i c_j c_k(-t) c_l(-t) a_\alpha^\dagger a_\beta^\dagger\rangle  
    &= 4 ( - (U U^\dagger)_{ij} (\hat U F \hat U^T)_{kl} + (U \hat U^\dagger)_{ik} ( U F \hat U^T)_{jl}  - (U\hat U^\dagger)_{il} (U F \hat U^T)_{jk} \\
    &\ \ \ \ \ \ - ( U F  U^T)_{ij} (\hat U \hat U^\dagger)_{kl} +( U F  \hat U^T)_{ik} (U \hat U^\dagger)_{jl} -( U F  \hat U^T)_{il} ( U \hat U^\dagger)_{jk} )
\end{aligned}
\end{equation}
\begin{equation}
\begin{aligned}
    J_3 & = \frac18 F_{\alpha \beta}F_{\gamma \delta}\langle c_i c_j c_k(-t) c_l(-t) a_\alpha^\dagger a_\beta^\dagger a_\gamma^\dagger a_\delta^\dagger\rangle
    \\
    &= 4( (UFU^T)_{ij} (\hat U F\hat U^T)_{kl} -(UF\hat  U^T)_{ik} (U F\hat U^T)_{jl}  + (UF\hat U^T)_{il} (U F\hat U^T)_{jk}
\end{aligned}
\end{equation}
\begin{equation}\label{eq:Jijkl}
\begin{aligned}
    \implies J_{ijkl} &= 4 \sqrt{\text{det} X}\left[ ( UU^\dagger - U F U^T)_{ij} (\hat U \hat U^\dagger - \hat U F \hat U^T)_{kl}\right. \\
    &- \left. ( U\hat U^\dagger - U F \hat U^T)_{ik} ( U \hat U^\dagger - U F \hat U^T)_{jl}
    + ( U\hat U^\dagger - U F \hat U^T)_{il} ( U \hat U^\dagger - U F \hat U^T)_{jk} \right]
\end{aligned}
\end{equation}
\end{widetext}
where all matrix multiplication $AB$ in these expressions \textit{is only over the first $N$ columns of $A$ the and first $N$ rows of $B$} even if $A$ or $B$ has dimension $2N \times 2N$.

Although this expression looks quite different from $I_{ij}$, if we were trying to evaluate the analogous expression, we would find
\begin{equation}
\begin{aligned}
    J_{ij} &= \langle c_i e^{-iH(M)}c_j \rangle = \sqrt{\text{det} X} \langle c_i c_j(-t) e^{\frac12 a_\alpha^\dagger F_{\alpha \beta} a_\beta^\dagger}\rangle\\
    &=2 \sqrt{\text{det} X}(U_{i\alpha}\hat U_{\alpha j}^\dagger  - (UF\hat U^T)_{ij}) 
      = 2 U_{i\alpha}I_{\alpha \beta}   U_{\beta j}^\dagger 
\end{aligned}
\end{equation}
which can be used to rewrite $J_{ijkl}$ accordingly. The last step follows making use of the unitarity of $\mathcal T$

If we are interested in computing similar quantities with more Majoranas, we can use Eq.~(C8) of \cite{udagawa2021} to prove that a modified Wick's theorem applies. This result explains why our Eq.~\eqref{eq:Jijkl} looks like it follows a Wick theorem with a different definition of a contraction.

\section{Convergence and other details from the numerics }\label{app:conv}

\begin{figure}[tbp!]
    \centering
    \includegraphics[width=0.45\textwidth]{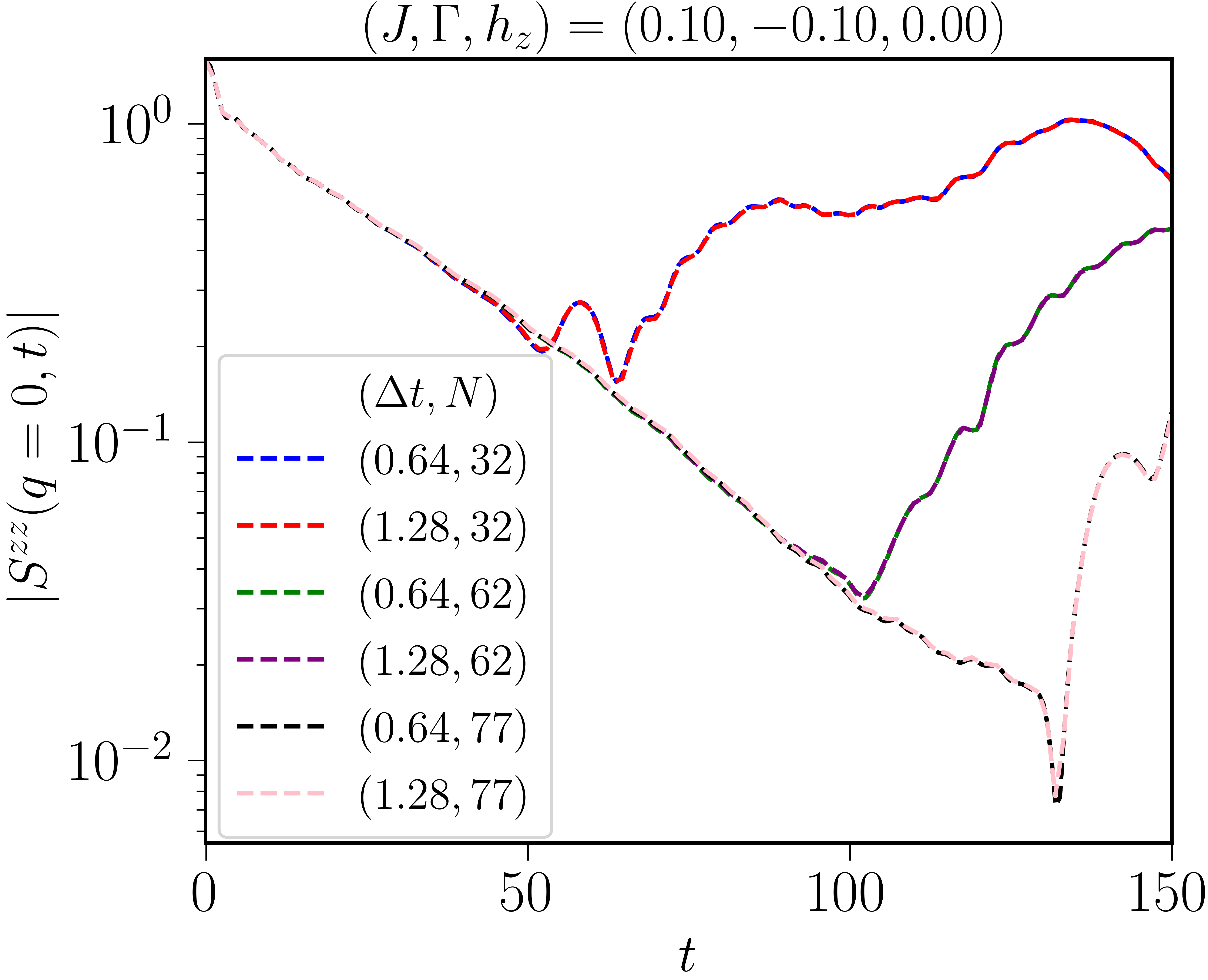}
    \caption{We plot $S^{zz}(q=0,t)$ for one parameter point but varying the time step $\Delta t$ and the number of sites, which is $2\times N \times N$. As argued in the main text, we see finite size effects at a time that roughly scales with $N$, which is easily identifiable as when the curve breaks off of a roughly exponential curve. }
    \label{fig:SQTconv}
\end{figure}

Since we are performing these calculations for finite systems in real space, the two main parameters that we should check convergence of are $N$, indicating the linear size of the system, and $\Delta t$, the time step after which we recompute the mean-field parameters and $S^{\alpha \beta}_{ij}(t)$. We plot a prototypical example in Fig.~\ref{fig:SQTconv}. As discussed in the main text, finite size effects appear at $t_c \sim N$, and this can roughly be seen as the curves break apart from the overall exponential decay.  

 We compute
\begin{equation}
    S^{\alpha \beta}(\pmb q,\omega)=\int_{-t_c}^{t_c} dt e^{i\omega t} S^{\alpha \beta}(\pmb q,t)
\end{equation}
where $S_{ij}^{\alpha \beta}(-t) = (S_{ji}^{\beta \alpha}(t))^*=(S_{ij}^{\beta \alpha}(t))^*$ where we take advantage of the translation and rotation symmetry. We estimate $t_c$ for each system size based on when the finite-size effects become clear. We only find slight differences for the largest $N$'s if we replace the abrupt cutoffs with a smooth one. 

We check convergence of $S^{\alpha \beta}(q,\omega)$ and find that $N\approx 80$ seems sufficient for all the parameter choices we make, except at the smallest $\omega$. We are limited from going to larger $N$, in general, because the time it takes to perform the largest system sizes and smallest time steps takes days to weeks, but, when $\Gamma\ne0$, memory also becomes a factor even though we are taking advantage of the reflection symmetry to reduce matrix size.

\section{A note about Gauge}\label{app:gauge}

With the transformation $S_i^\alpha=i c_i b_i^\alpha $, the Hilbert space has been expanded, so, properly, we should project the wave-function we obtain back into the physical Hilbert space \cite{kitaev2006}.  The projection opertaor has the form
\begin{equation}
    P = \prod_i \frac{1+D_i}2
\end{equation}
where $D_i = c_i b_i^x b_i^yb_i^z$. The projection operator commutes with all the spin operators $S_i^\alpha$ and therefore also the Hamiltonian. Additionally, $P^2=P$, as should be expected. 

In applying mean-field theory, many works handle the projection by imposing the constraint on average \cite{Jiang2020, Choi2018, Zhang2021, Okamoto2013, Seifert2018, Ralko2020,Gao2019}, arguing that the effect is higher-order \cite{Go2019}, using a different transformation without a gauge issue \cite{Nasu2018static, Nasu2019}, or ignoring the effect altogether \cite{Mei2012,li2022,Schaffer2012}. In our formalism, in zero-field, we automatically satisfy the constraints, on average, as expressed in \cite{Ralko2020}. 

To fully take account of the gauge, we should alternatively compute
\begin{equation}
    S_{ij}^{\alpha \beta} = \frac14 e^{iE_{MF}t} \frac{\langle v|P c_i b_i^\alpha U(t,0)c_j b_j^\beta\rangle }{\langle P \rangle}.
\end{equation}
If we imagine expanding out $P$, we need to consider the contribution from many different terms with various numbers of $D_i$. Focusing only on the exact case, every term with at least one $D_i$ must vanish as is evident from rewriting the $b_i^\alpha$'s in terms of bond fermions \cite{Baskaran2007}. The only exception is the term with all $D_i$ does not vanish by this argument. However $\mathcal D = \prod_i D_i \sim \prod_i c_i$ since all the $b_i^\alpha$ pair up into the conserved quantities $u_{\langle ij\rangle_\alpha}=ib_i^\alpha b_j^\alpha$. The operator $\prod_i c_i$ commutes with the Hamiltonian and the $u_{\langle ij \rangle_\alpha}$. Ignoring the complications from having a gapless point, we can see then that $\prod_i D_i$ acting on the ground state just gives a constant. 
Beyond the exact point, the Hamiltonian still commutes with $\mathcal D$, which implies that we can group any term, $\alpha$, in the expansion of $P$, with the term $\alpha D$ to just get an overall prefactor $1+\langle \mathcal D\rangle$ provided we limit which terms we consider accordingly. We expect terms with fewer than all the $D_i$ to be suppressed by correlations that are small. Limiting our analysis the the zero-field case, we need an even number of $D_i$ to have the correct number of $c_i$. The analysis of which terms are most important is complicated because, there are $2N_x N_y \choose m$ terms with products of $m$ $D_i$. A reasonable guess, though, would be that the leading order correction to our expression in the main paper would be from the terms with the fewest numbers of $D_i$. Namely
\begin{equation}
  S_{ij}^{\alpha \beta} = \frac14 e^{iE_{MF}t} \frac{\langle v|(1+\sum_{k,l>k} D_k D_l)  c_i b_i^\alpha U(t,0)c_j b_j^\beta\rangle }{1+ \sum_{k,l>k}\langle D_k D_l\rangle  }.
\end{equation}
However, we find that $\sum_{k,l>k}\langle D_k D_l\rangle$ scales linearly with the number of sites implying that such a term might provide a large correction in the thermodynamic limit even for small perturbations.

If we are interested in the case where $J=\Gamma=h_y=h_x=0$ and $h_z\ne 0$, we can use the Jordan-Wigner formalism \cite{Nasu2018static}. In this case we just need to compute $S_{ij}^{zz}(t)$, which is exact, and $S_{ij}^{xx}(t)$, which will contain Jordan-Wigner strings. By picking site $j$ to be the site where $S^x_j = a_j/2$ (i.e. the unique site without a string operator), and using periodic boundary conditions, the expression for $S_{jj}^{xx}(t)$ is equivalent to our approach above. The ``flipping'' of the sign of the $a_j b_{j+x}$ term occurs because it is scaled by $P_{F,0}$, the string operator containing the product of all the $(-2S_j^z)$ in the first ``row'' of the honeycomb lattice (all the sites connected just by $x$ and $y$ bonds), which changes sign upon the operator of $a_j=2S_j^x$. Additionally, this operator $P_{F,0}$ commutes with the Hamiltonian and has a value of $1$ in the ground state, which implies that $S_{j+x,j}^{xx}(t)$ also receives no correction. However, terms like $S_{j+y,j}^{xx}(t)$ and $S_{j+z,j}^{xx}(t)$ do receive corrections, which could be systematically included, but should be suppressed by a factor of $h_z/|K|$. 

To summarize, our approach handles the projection operator similarly to other works in the literature,  and we provide a potential path to include the neglected effects. It would be beneficial, in future work, to quantify the errors that these approximations produce. 

\bibliography{KST.bib}

\end{document}